\documentclass[aps,prd,12pt]{revtex4}
\usepackage{graphics,amssymb,epsfig,float,psfrag,axodraw}
\usepackage[usenames,dvips]{color}
\usepackage{rotating}

\begin{document}
\renewcommand{\baselinestretch}{1.0}
\newcommand{\lsim}{\raisebox{-0.13cm}{~\shortstack{$<$ \\[-0.07cm] $\sim$}}~}
\newcommand{\gsim}{\raisebox{-0.13cm}{~\shortstack{$>$ \\[-0.07cm] $\sim$}}~}
\newcommand{\lra}{\longrightarrow}
\newcommand{\ee}{e^+e^-}
\newcommand{\gam}{\gamma \gamma}
\newcommand{\s}{\smallskip}
\newcommand{\nn}{\noindent}
\newcommand{\non}{\nonumber}
\newcommand{\beq}{\begin{eqnarray}}
\newcommand{\eeq}{\end{eqnarray}}
\newcommand{\pslash}{\not\hspace*{-1.6mm}p}
\newcommand{\kslash}{\not\hspace*{-1.6mm}k}
\newcommand{\lslash}{\not\hspace*{-1.6mm}l}
\newcommand{\eslash}{\hspace*{-1.4mm}\not\hspace*{-1.6mm}E}

\title{On Higgs mass generation mechanism in the Standard Model}

\author{V.A.~Bednyakov}
\affiliation{Dzhelepov Laboratory of Nuclear Problems,
         Joint Institute for Nuclear Research, \\
         141980 Dubna, Russia; E-mail: Vadim.Bednyakov@jinr.ru}
\author{N.D.~Giokaris}
\affiliation{Physics Department, The University of Athens}
\author{A.V.~Bednyakov}
\affiliation{Bogoljubov Laboratory of Theoretical Physics, JINR}


\begin{abstract} 
      The mass-generation mechanism is 
      the most urgent problem of the modern particle physics.      
      The discovery and study of the Higgs boson with
      the Large Hadron Collider at CERN
      are the highest priority steps to solve the problem. 
      In this paper, the Standard Model 
      Higgs mechanism of 
      the elementary particle mass generation 
      is reviewed with pedagogical details. 
      The discussion of the Higgs quadric self-coupling  
      $\lambda$ parameter and the 
      bounds to the Higgs boson mass are presented.
      In particular, the unitarity, triviality, and stability 
      constraints on the Higgs boson mass are discussed.
      The generation of the finite value for the $\lambda$ parameter 
      due to quantum 
      corrections via effective potential is illustrated.
      Some simple predictions for the top quark 
      and the Higgs boson masses 
      are given when both the top Yukawa coupling and 
      the Higgs self-coupling $\lambda$ are equal to 1.

\end{abstract}
\maketitle

\section{Introduction}
    One of the highest priorities of particle physics today is the 
    discovery of the dynamics responsible for Electro-Weak Symmetry 
    Breaking (EWSB)
\cite{Haber:2004tm}.
    In the Standard Model, nowadays the main working paradigm 
    of particle theory, this dynamics is expected due to 
    self-interactions of special complex scalar fields.  
    This approach predicts the existence of one physical
    elementary scalar, the so-called Higgs boson 
\cite{Gunion:1989we,Gunion:1992hs}. 
    A search for and the discovery of this still-escaping 
    boson, and investigation of its properties 
    are practical steps to solve the problem of EWSB
    which are currently planned to perform with 
    the Large Hadron Collider (LHC) 
    and in future with the International Linear Collider (ILC). 

    The modern Standard Model (SM) of particle physics 
    is a unified framework to describe 
    electromagnetic and weak interactions between quarks and leptons 
    together with strong interactions between quarks
(see, for example, \cite{Djouadi:2005gi}).
    It is the Yang-Mills theory based on the electroweak symmetry group
    SU(2)$_{\rm L}\times $U(1)$_{\rm Y}$ of Glashow, Weinberg, and Salam 
\cite{Glashow:1961tr,Weinberg:1967tq,Salam:1969aa} 
    and strong SU(3)$_{\rm C}$ group of QCD
\cite{Gell-Mann:1964nj,Fritzsch:1973pi,Gross:1973id,Politzer:1973fx}.
    This model (before the electroweak symmetry breaking) has only 
    matter and gauge fields. 
    The matter 
    fields are composed of three generations of fermions (spin-1/2),  
    left-handed and right-handed quarks and leptons, 
    $\displaystyle f_{\rm L,R} =\frac{1 \mp \gamma_5}{2}f$. 
    It is {\em crucial}\/ for our consideration that
    the left-handed fermions are in the weak SU(2)$_{\rm L}$ 
    isodoublets, while the right-handed fermions are 
    weak isosinglets.
    Moreover, both left- and right-handed quarks are triplets under 
    the ${\rm SU(3)_C}$ group, while all leptons are color singlets. 
    The gauge fields mediate the above-mentioned interactions 
    and correspond to the (spin-1) bosons.
    In the electroweak sector, the field 
    $B_\mu$ corresponds to  the U(1)$_{\rm Y}$ group 
    and the three fields $W^{1,2,3}_\mu$ correspond to 
    the SU(2)$_{\rm L}$ group.
    There is also an octet of gluon fields 
    $G_\mu^{a}$ which correspond to  the color ${\rm SU(3)_C}$ group.
    Due to the non-Abelian nature of the SU(2) and SU(3) groups, 
    there are triple
    and quartic 
    self-interactions between their gauge fields
    $V_\mu = W_\mu $ or $G_\mu$.
     The matter fields $\psi$ are minimally coupled to the gauge 
     fields through the covariant derivative $D_\mu$
(see Appendix~\ref{App-SM}),
     which leads to a unique form of interaction 
     between the fermion and gauge fields, 
     ($-g_i \overline \psi V_\mu \gamma^\mu \psi$),
    where 
    $g_s$, $g_2$, and $g_1$ are, respectively, 
    the coupling constants of 
    ${\rm SU(3)_C}$,  ${\rm SU(2)_L}$, and ${\rm U(1)_Y}$.  

      The SM Lagrangian, without mass terms for fermions and 
      gauge bosons, is then given by 
\begin{eqnarray}
\label{SM-Lagrangian}
{\cal L}_{\rm SM}&=& -\frac{1}{4} G_{\mu \nu}^a G^{\mu \nu}_a 
-\frac{1}{4} W_{\mu \nu}^a W^{\mu \nu}_a -\frac{1}{4} 
B_{\mu \nu}B^{\mu \nu} \\ 
&& + \bar{L_i}\, i D_\mu \gamma^\mu \, L_i + \bar{e}_{Ri} \, i D_\mu 
\gamma^\mu \, e_{R_i} \ 
+ \bar{Q_i}\, i D_\mu \gamma^\mu \, Q_i + \bar{u}_{Ri} \, i D_\mu 
\gamma^\mu \, u_{R_i} \ + \bar{d}_{Ri} \, i D_\mu \gamma^\mu \, d_{R_i}. \non 
\end{eqnarray}
        This Lagrangian is invariant under local 
	${\rm SU(3)_C \times SU(2)_L \times U(1)_Y}$ gauge 
	transformations for fermion and gauge fields.
	Here $L_i$ and $Q_i$ denote the left-handed lepton and quark
	doublets, respectively,  
	while $f_R$ denotes the relevant right-handed singlets.
	In the case of the electroweak sector, for instance, one has  
	the gauge transformations:
\begin{eqnarray}
L(x) \to L'(x)=e^{i\alpha_a(x) T^a + i \beta(x)Y } L(x) \ \ , \ \
R(x) \to R'(x)=e^{i \beta (x) Y} R(x) \non \\
\vec{W}_\mu (x) \to \vec{W_\mu}(x) -\frac{1}{g_2} 
\partial_\mu \vec{\alpha}(x)- 
\vec{\alpha}(x) \times \vec{W}_\mu(x) \ , \ B_\mu(x) \to B_\mu(x) -  \frac{1}
{g_1} \partial_\mu \beta (x) .
\end{eqnarray}
       The gauge fields and the fermion fields are {massless} here.
       More details one can find in Appendix~\ref{App-SM}.

       It is interesting to note 
       that in the case of strong interactions
       (while the gluons are indeed massless particles) 
       the mass terms of the form  
       $-m_q\overline{\psi}\psi$ can be generated for
       the colored quarks in an SU(3) gauge invariant way.
       This is due to the fact that all (left- and right-handed) 
       quarks belong only to triplets of the SU(3) color group
       and all transform in the same manner.

       On the contrary, 
       the situation in the electroweak sector is really horrible.
       Indeed, one knows experimentally 
       that the weak gauge bosons are massive and the weak 
       interaction is very short ranged.
       However, as soon as one adds standard mass terms for the gauge bosons, 
       $\frac{1}{2} M_W^2 W_\mu W^\mu$,
       one immediately violates the 
       local SU(2)$\times$U(1) gauge invariance.
       This is clearly seen for the QED where the photon 
       is massless because of the U(1) local gauge symmetry.
       Indeed, the transfromed ``photon'' mass term 
\begin{eqnarray}
\frac{1}{2}M_\gamma^2 A_\mu A^\mu \to \frac{1}{2}M_\gamma^2 
(A_\mu - \frac{1}{e} 
\partial_\mu \alpha) (A^\mu - \frac{1}{e} \partial^\mu \alpha) \neq
\frac{1}{2}M_\gamma^2 
A_\mu A^\mu
\end{eqnarray} 
       can hold its form untouched only if $M_\gamma^2 \equiv 0$.
       In addition, if one includes explicitly the mass term 
       $-m_f \overline{\psi}_f \psi_f$ for the
       SM fermion $f$ in the Lagrangian, 
       then, for instance, one would have for the electron 
\begin{eqnarray}\label{Intro-m_e}
- m_e \bar{e}e \equiv -m_e \bar{e} \bigg( \frac{1-\gamma_5}{2}
+\frac{1+\gamma_5}{2}\bigg) e= -m_e(\bar{e}^{}_R e^{}_L+\bar{e}^{}_Le^{}_R) 
\end{eqnarray}
       which is obviously noninvariant under the weak isospin symmetry 
       transformations discussed above, since $e^{}_L$ is a member 
       of the SU(2)$_{\rm L}$ doublet, while 
       $e_R$ is the SU(2)$_{\rm L}$ singlet and, therefore, they change 
       under transformation in a different manner.

       Therefore, the mass terms for gauge bosons and 
       fermions induced ``by-hand'' lead to an obvious breakdown 
       of the local ${\rm SU(2)_L\times U(1)_Y}$ gauge invariance. 
       The unbroken symmetry means that all 
       fundamental particles have to be {\em massless}.\/
       This is because both the fermion mass term $f_L\times  f_R$
       and that of gauge bosons are not SU(2)$_L$ invariant
\cite{Kazakov:1989ny}.
       One can see, that generation of the mass for an elementary particle 
       in the SM is strongly connected with the symmetry violation.
       One needs a mechanism for this violation, and one believes, 
       that this mechanism
       will simultaneously allow the elementary particles to obtian
       their masses.

       In principle, the idea of mass generation due to interaction
       is rather simple.
       Consider the renormalizable Lagrangian ${\cal L}=g A_\mu A_\mu\phi$
       describing interaction of the scalar field $\phi$ with
       the massless vector field $A_\mu$.  
       In an ordinary theory mean vacuum expectation values (vev's) 
       are zero.
       Assume now that the scalar field has nonvanishing vev  
       $v\ne 0$, so 
       $\phi = v + \sigma$ with $\langle 0| \sigma| 0 \rangle =0$.
       The Lagrangian becomes 
       ${\cal L} = g v A_\mu A_\mu  + g A_\mu A_\mu \sigma$.
       The first term is a right mass term and the
       vector particle obtains a mass $m^2= 2 v g$. 
       {\em The only question is where $v\ne 0$ comes from?}
\cite{Gaillard:1977wu}. 
       In other words, 
       is there a way to generate the gauge boson and
       the fermion masses without violating  
       SU(2)$\times$U(1) gauge invariance? 
       The positive answer is given by 
       Higgs, Kibble and others
\cite{Higgs:1964pj,Higgs:1966ev,Kibble:1967sv}.
       This is the
       spontaneous symmetry breaking Higgs mechanism 
(see \cite{Djouadi:2005gi} and Appendix \ref{App-SM}).

       In fact, 
       the Higgs mechanism is needed due to the 
       SU(2)$\times$U(1) gauge structure of the SM.
       It is remarkable that from the practical point of view 
       the mass generation 
       by means of the Higgs mechanism 
       in the SM is forced by the $V$--$A$ 
       structure of the weak interaction 
       (and in some sense by the  
       absence of the right-handed neutrinos $\nu_R$\/
       or even by the masslessness of all neutrinos). 
       
       Below, in discussing the Higgs mechanism and related topics
       we follow, to a large extent, the excellent review 
       of A.~Djouadi
\cite{Djouadi:2005gi}. 

\section{Higgs Mechanisms}
\subsection{The simplest example} 
\label{simplest}
      First of all,   
      consider a simple Lagrangian for 
      a scalar real field $\phi$ 
\begin{eqnarray}\label{Lagr-scalar}
{\cal L}= \frac12 \partial_\mu \phi \, \partial^\mu \phi - V(\phi), 
\qquad {\rm where} \qquad 
V(\phi)= \frac{1}{2} \mu^2 \phi^2 + \frac{1}{4}\lambda \phi^4.
\end{eqnarray}
      Since the potential should be bounded from below,
      the self-coupling $\lambda>0$. 
      With the mass term $\mu^2>0$, 
      the potential $V(\phi)$ is always positive.
      Furthermore, the $\phi^4$-term  
      describes self-interaction with intensity $\lambda$.
      Other terms $\phi^n$ with $n>4$ have to be excluded from
      consideration because they produce infinities in
      calculated observables
\cite{Kane:1987gb}. 
      The case when the potential $V(\phi)$ also contains an extra 
      $\phi^3$-term is considered in Appendix 
\ref{App-Phi3}.  

\begin{figure}[!ht] 
\begin{center}
\vspace*{-2.cm}
\hspace*{-2cm}
\epsfig{file=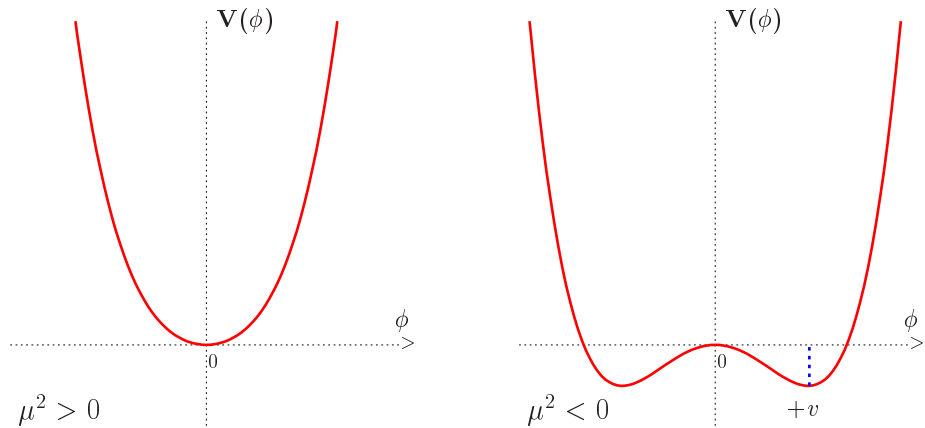,width=14.cm} 
\end{center}
\vspace*{-12.3cm}
\caption{The potential $V$ of the scalar field $\phi$ in 
         the case $\mu^2>0$ (left) and $\mu^2 <0$ (right).
	 From
\cite{Djouadi:2005gi}.
\label{V-Higgs}
} \vspace*{-3mm}
\end{figure}

    To find an excitation spectrum of the system described by Lagrangian
(\ref{Lagr-scalar}),
    one first has to find minimum (or minima) of the potential $V(\phi)$. 
    The system has minimal energy when its 
    both kinetic and potential energies separately are minimal. 
    The kinetic energy is minimal when $\phi$ is a constant.
    The minimum gives one a classical main (vacuum) state of the system.
    Next, one has to decompose the field $\phi$ in the vicinity of
    this main state and has to find excitation states.
    In a field theory the main state is the vacuum and the
    excitations are particles.
    Particle masses are defined by the form of
    the Lagrangian in the vicinity of the classical minimum
\cite{Kane:1987gb}. 
      When $\mu^2>0$ (left panel in 
Fig.~\ref{V-Higgs}),
      the minimum of potential 
(\ref{Lagr-scalar}) 
      is reached at $\phi=0$. 
      Therefore, vacuum expectation value for the field 
      $\langle 0| \phi | 0 \rangle \equiv \phi_0=0$. 
      Lagrangian 
(\ref{Lagr-scalar}) 
       then simply describes a spin-zero particle of mass $\mu$.
      It is also invariant under the reflexion symmetry 
      $\phi \to -\phi$ since there are no cubic terms.  

     If $\mu^2 <0$ (right panel in Fig.~\ref{V-Higgs}), 
     the potential $V(\phi)$ has minima
     not at $\phi_0=0$ but at $\phi_1$ and $\phi_2$  
     which solve the minimum condition \
     $\partial V/\partial \phi = \phi(\mu^2 +\lambda \phi^2)=0$.
     Now the system has two states (vacuums) with the lowerst energy 
     $\displaystyle V_{\min}= - \frac{v^4\lambda}{4}< 0$ \
     at 
\begin{eqnarray}
\phi_1 = \sqrt{-\frac{\mu^2}{\lambda}} \equiv v > 0 , 
\quad {\rm and} \quad
\phi_2 = -\sqrt{-\frac{\mu^2}{\lambda}} \equiv - v.
\end{eqnarray}  
     The quantities 
     $\phi_1 \equiv + v$ and
     $\phi_2 \equiv - v$ are
     the vacuum mean values of the field $\phi$ and are also
     called the vacuum expectation value (vev) of the scalar field $\phi$. 
     Lagrangian 
(\ref{Lagr-scalar}) 
     no longer describes a particle with mass $\mu$. 

    To find now energies of the particles 
    (and to interpret correctly the theory), one has to 
    choose one of the minimum, e.g., with $\phi=v$,
    and investigate the situation in 
    the vicinity of the minimum  
    of the potential $V(\phi)$.
    To this end, one 
    introduces a new scalar field $\sigma$ in such a way
    that $\phi= v + \sigma$ and 
      $\langle 0| \sigma |0\rangle = 0$.
    Furthermore, 
    one has to expand all the terms in Lagrangian
(\ref{Lagr-scalar}) 
    in series in the small parameter
    $\sigma$ around the potential minimum at $\sigma=0$.  
    In terms of the new field $\sigma$, the Lagrangian becomes 
\begin{eqnarray*}\nonumber
{\cal L} &=& 
            \frac{1}{2}\partial_\mu\sigma\, \partial^\mu\sigma 
            -\left\{
	    \frac{\mu^2}{2}[v^2+2v\sigma+\sigma^2]
	    +\frac{\lambda}{4}
	    [v^4+4v^3\sigma+6v^2\sigma^2+4v\sigma^3 + \sigma^4]
	    \right\} = \\
	 &=&
            \frac{1}{2}\partial_\mu\sigma\,\partial^\mu\sigma 
           -\left\{
	     \frac{v^2}{2}(\mu^2+\frac{\lambda v^2}{2})
	    + v\sigma (\mu^2 + \lambda\sigma^2) 
	    + \frac{\sigma^2}{2}(\mu^2 + 3v^2\lambda)
            + \lambda v \sigma^3
	    + \frac{\lambda}{4}\sigma^4
	    \right\}.
\end{eqnarray*}
       With the minimum relation $\mu^2=-\lambda v^2$ 
       the linear term disappears and one finally has 
\begin{eqnarray}\label{Lagr-scalar-sigma}
{\cal L}
       &=& \frac{1}{2} \partial_\mu \sigma \, \partial^\mu \sigma 
          - \frac{2\lambda v^2}{2} \sigma^2 
	  - \lambda v \sigma^3 
	  - \frac{\lambda}{4} \sigma^4 
	  + \frac{\lambda v^4}{4}. 
\end{eqnarray}
      Due to the correct sign of the $\sigma^2$-term one can 
      interpret it as a mass term, thus Lagrangian
(\ref{Lagr-scalar-sigma}) describes a scalar field of 
      mass $m_{\sigma}^2=2\lambda v^2= -2\mu^2$, 
      with $\sigma^3$ and $\sigma^4$ being self-interactions.
      The new mass $m_{\sigma}$ was generated due to  
      self-interactions of the field $\sigma$
\cite{Kane:1987gb},    
      and $m^2_\sigma>|\mu^2|$ means that 
      the back-attractive ``force'' for the
      new $\sigma$ field would be stronger than for the 
      initial field $\phi$. 
      Since there are now cubic terms, the 
      reflexion symmetry is broken.
      This is the simplest example of spontaneously broken symmetry.
      The symmetry is violated by means of 
      inevitable alternative --- one must  
      choose only one concrete  
      vacuum (at $\phi_1=v$, or at $\phi_2=-v$). 
      After that
      the unique 
      vacuum does not possess the symmetry of the initial Lagrangian
      (\ref{Lagr-scalar}). 
	Actually, the symmetry transformation turns one vacuum state
	(with $\phi_1=v$)  
	into the other one (with $\phi_2=-v$).

      The Lagrangian 
(\ref{Lagr-scalar-sigma}) now has the potential with 
      {\em nonzero cubic term},  
$$V(\sigma) = \lambda v^2\sigma^2 +\lambda v\sigma^3 
	    + \frac{\lambda}{4} \sigma^4-\frac{\lambda v^4}{4}.
$$
      Due to this term the potential {\em could},\/ in principle,
      have minimum at $\sigma\ne 0$, 
      which spoils the main condition 
      $\langle 0| \sigma |0\rangle = 0$.
      (See the discussion of the Higgs mechanism with the extra 
      $\phi^3$-term in Appendix 
\ref{App-Phi3}.)
     Applying the extremum condition to this potential 
     one has the relation 
\begin{equation}\label{Sigma-minimum}
\frac{\partial V}{\partial \sigma}
	       = \lambda \sigma (\sigma^2 + 3 v \sigma+ 2 v^2 ) 
               = \lambda \sigma (\sigma + v) (\sigma + 2v) = 0.
\end{equation}
      There are 3 extrema with 
      $V(\sigma=0)=-{\lambda v^4}/{4} < 0$, \ 
      $V(\sigma=-v)= \lambda v^4 (1-1+1/4-1/4) =0$,  and 
      $V(\sigma=-2v) 
      = \lambda v^4 (4 - 8 + 16/4 - 1/4) =  
      V(\sigma=0)=-{\lambda v^4}/{4}$.
      Therefore, two minima have the same depth and
      we can safely choose as a true vacuum the minimum at
      $\sigma=0$ which indeed has $\langle 0| \sigma |0\rangle = 0$. 
      In fact, it is not surprising that the cubic term does not spoil
      the vacuum. 
      Moving into only one vacuum state one physically 
      does not introduce any new dynamics. 
      Therefore, there is no reason to change 
      the shape of the potential and it remains unchanged.
      However, 
      to reproduce the unchanged shape of the potential
      in the new (shifted) coordinate framework 
      (where $\langle 0| \sigma |0\rangle = 0$), 
      one needs right this cubic term.
     
\subsection{The Higgs mechanism with a complex scalar field}
\label{HM-CSF}
    The relevant Higgs Lagrangian for a complex scalar field 
    $ \phi = \frac{1}{\sqrt{2}}(\phi_1 + i\phi_2)$ is
\begin{equation}
\label{complex-1}
{\cal L} = 
  \left({\partial_\mu\phi}\right)^* \left({\partial^\mu\phi}\right) 
    -\mu ^2\phi^*\phi - \lambda\left({\phi^*\phi}\right)^2
\end{equation}
      with at least $\lambda>0$. 
      This Lagrangian is invariant under global gauge 
      transformations $\phi \to \phi'=e^{i\chi}\phi$ 
      and, therefore, has global U(1)-symmetry.
      When $\mu ^2 < 0$ the scalar potential 
\begin{eqnarray*}
V(\phi) &=& \mu^2  \phi^* \phi + \lambda (\phi^* \phi)^2
	   =  \frac{\mu^2}{2} (\phi_1 - i\phi_2)  
	                       (\phi_1 + i\phi_2)
	   + \frac{\lambda}{4}((\phi_1 - i\phi_2) 
	                       (\phi_1 + i\phi_2))^2= \\
           &=& \frac{\mu^2}{2}(\phi^2_1+\phi^2_2) 
	      + \frac{\lambda}{4}(\phi^2_1+\phi^2_2)^2 
\end{eqnarray*}
     has minimum values of 
$\displaystyle 
V(\phi)_{\min} 
= \frac{\mu^2}{2}v^2 + \frac{\lambda}{4} v^4
      =-\frac{\lambda}{4}v^4 < 0 \
$
     at $\displaystyle \phi^2_0= \frac{v^2}{2}$ 
     along a circle of the radius $v$ 
     in the ($\phi_1,\phi_2$), plane where $v$ is given by 
\begin{equation}
\label{complex-2}
v^2 = \phi^2_1+\phi^2_2, \qquad
v^2 = \frac{-\mu^2}{\lambda}> 0.
\end{equation}
     To construct a theory, one has to investigate the situation
     in the vicinity of one of the minima in the circle.     
     To this end, one has to choose one of the minima 
     (to violate the symmetry of all possible solutions). 
     One can take 
     the real scalar field $\phi_1$ with the nonzero vacuum
     expectation value $\phi_1=v$, while the imaginary one $\phi_2=0$
     at the minimum.
     Furthermore,   
     the scalar complex field $\phi$ can be parameterized also
     in the form (with both real $\eta(x)$ and $\xi(x)$)
$$
\phi(x) = \frac{1}{\sqrt{2}}\left(v+\eta(x)+i\xi(x)\right)
$$
     with $\eta(x)=\xi(x)=0$ at $V(\phi)_{\min}$.
     Therefore, after introduction of  a Higgs mass
\begin{equation}\label{complex-H}
M_h = \sqrt{2 \lambda v^2} \equiv \sqrt{2}|\mu|
\end{equation} 
     Lagrangian 
(\ref{complex-1}) has the form (see Appendix
\ref{App-proves}): 
\begin{eqnarray}\label{complex-5}
{\cal L}
     &=& 
      \frac{1}{2}\partial_\mu\eta\,\partial^\mu\eta
     +\frac{1}{2}\partial_\mu\xi  \,\partial^\mu\xi
     - \frac{M^2_h}{2}\eta^2 
     -\frac{\lambda}{4}(\eta^2+\xi^2)^2
     - v\lambda\eta^3 - v\lambda\eta\xi^2 + \frac{v^4\lambda}{4}.
\end{eqnarray}
        Now this Lagrangian describes interaction
	between two real scalar fields $\eta(x)$ and $\xi(x)$ 
	(both with zero vev's).
	The $\eta(x)$ (Higgs) field 
	is massive with the mass given by 
(\ref{complex-H}) and $\xi(x)$ is massless.  
        The physical reason is the following.
	Radial excitations (described by $\eta$)
	are against the increase of the potential.
	The potential forces the relevant particles 
	to go back to the minimum and 
	these excitations are massive.
	Excitations in the direction of the circle
	have locally no any back force at all 
	and these excitations are massless.  
	This is the first example of the Goldstone theorem
	(when global symmetry 
	is spontaneously broken, the massless boson appears),
	which we consider below in a bit more detail. 

     Consider, following A.Djouadi 
\cite{Djouadi:2005gi},
     four real scalar fields $\phi_i$ with 
     $i=0,1,2,3$ with a Lagrangian (the summation over the
     index $i$ is understood)
\begin{eqnarray}
{\cal L}= \frac12 \partial_\mu \phi_i \, \partial^\mu \phi_i 
        - \frac{\mu^2}{2} (\phi_i \phi_i)
        - \frac{\lambda}{4} (\phi_i \phi_i)^2 
\end{eqnarray}
   which is invariant under the rotation group in four dimensions 
   O$(4)$, $\phi_i(x) = R_{ij} \phi_j (x)$ for any orthogonal matrix $R$. 
   Again, for $\mu^2<0$, the potential has a minimum at 
   $\phi_i^2 = - \mu^2/ \lambda \equiv v^2$ where $v$ is the vev. 
   As previously, we expand around one of the minima, 
   $\phi_0= v+ \sigma$, and rewrite the fields $\phi_i=\pi_i$ 
   with $i=1,2,3$. 
   The Lagrangian in terms of the 
   new fields $\sigma$ and $\pi_i$ becomes
\begin{eqnarray}
{\cal L} &=& \frac{1}{2} \partial_\mu \sigma \, \partial^\mu \sigma - 
\frac{1}{2} (- 2\mu^2) \sigma^2 - \lambda v \, \sigma^3- \frac{\lambda}{4} 
\, \sigma^4  \non \\ 
&+ &  \frac{1}{2} \partial_\mu \pi_i \, \partial^\mu \pi_i  
- \frac{\lambda}{4} (\pi_i \pi_i)^2 -  \lambda v \pi_i \pi_i 
\sigma -\frac{\lambda}{2} \pi_i \pi_i \sigma^2 .
\end{eqnarray}
        As expected, we still have a massive $\sigma$ boson with 
	$m^2=-2\mu^2$, 
	but also we have three massless ``pions'', since now 
	all the bilinear $\pi_i \pi_i$ 
	terms in the Lagrangian have vanished. 
	Note that there is still O(3) 
	symmetry among the $\pi_i$ fields.

	This brings us to state {\em the Goldstone theorem}
\cite{Goldstone:1961eq,Goldstone:1962es}: 
       For every spontaneously broken continuous symmetry, 
       the theory contains  massless scalar (spin-0) particles 
       called Goldstone bosons. The number of Goldstone bosons is
       equal to the number  of broken generators. 
       For O$(N)$ continuous symmetry, there are 
       $\frac{1} {2}N(N-1)$ generators;
       the residual unbroken symmetry O$(N-1)$ has 
       $\frac{1}{2} (N-1)(N-2)$ generators
       and, therefore, there are $N-1$ massless Goldstone bosons, 
       i.e. 3 for the O$(4)$ group. 

\subsection{The Higgs mechanism in an Abelian theory} 
\label{HM-CSF_A}
      A rather simple case of local Abelian U(1) symmetry
      contains a complex scalar field and 
      an electromagnetic field $A_\mu$ 
\begin{eqnarray}\label{local-U1}
{\cal L} = - \frac{1}{4} F_{\mu \nu} F^{\mu \nu} 
           + (D_\mu \phi)^* (D^\mu \phi) - V (\phi) 
\end{eqnarray}
     with the covariant derivative 
     $D_\mu= \partial_\mu - ie A_\mu$ and with the scalar potential 
     (see also 
\cite{Gaillard:1977wu}) 
\begin{eqnarray}
          V(\phi)=  \mu^2 \phi^* \phi + \lambda \ (\phi^* \phi)^2
	         = \mu^2 |\phi|^2 + \lambda |\phi|^4.  
\end{eqnarray}
    The Lagrangian 
(\ref{local-U1}) is renormalizable and
    invariant under the local gauge U(1) transformation
\begin{eqnarray} \label{Ab-gauge}
      \phi(x) \to e^{i \alpha(x)} \phi(x), \qquad 
      \phi(x)^{\dag} \to e^{-i \alpha(x)} \phi(x)^{\dag}, \qquad 
      A_\mu (x) \to A_\mu (x) + \frac{1}{e} \partial_\mu \alpha(x). 
\end{eqnarray} 
      The local gauge invariance demands introduction of the
      massless vector field $A_\mu$
\cite{Kane:1987gb}.  
      For $\mu^2>0$, Lagrangian 
(\ref{local-U1}) is the QED Lagrangian for a charged scalar 
      particle of mass $\mu$ and with $\phi^4$ self-interactions. 
      For $\mu^2<0$ the field $\phi(x)$ will acquire a 
      vacuum expectation value 
      and the minimum of the potential $V(\phi)$ will be at
\begin{eqnarray}
     \langle\phi\rangle_0 \equiv \langle 0|\phi|0\rangle 
      = \left(-\frac{\mu^2}{2\lambda} \right)^{1/2} 
      \equiv\frac{v}{\sqrt{2}} .
\end{eqnarray}
      We expand the Lagrangian around the 
      vacuum state $\langle \phi \rangle_0$ 
\begin{eqnarray} \label{Ab-phi-v}
       \phi(x) = \frac{1}{\sqrt{2}}\left(v+\eta(x)+i\xi(x)\right)
\end{eqnarray}
      and assuming that $\langle 0|\eta|0 \rangle = 
      \langle 0| \xi |0\rangle = 0$. 
      With 
(\ref{Ab-phi-v}) Lagrangian 
(\ref{local-U1}) becomes (see Appendix
 \ref{App-proves}): 
\begin{eqnarray} \label{Abelian-Lagrangian-1} 
{\cal L} 
	&=&
         \frac{v^4\lambda}{4} 
	-\frac{1}{4}F_{\mu \nu}F^{\mu \nu} 
            +\frac{e^2v^2}{2} A_\mu A^\mu 
            +\frac{1}{2}\partial_\mu\eta\,\partial^\mu\eta
	    -\frac{2v^2\lambda}{2}\eta^2 
            +\frac{1}{2}\partial_\mu\xi \,\partial ^\mu\xi 
	   - {ev} A^\mu \partial_\mu\xi 
\\&+& \nonumber 
      \frac{e^2}{2} A_\mu A^\mu (2 v \eta +\eta^2 + \xi^2)
       +  e A^\mu \xi\partial_\mu\eta 
       - e A^\mu \eta\partial_\mu\xi
       - \frac{\lambda}{4}(\eta^2 + \xi^2)^2
       - {v\lambda}\eta (\eta^2 +\xi^2) .
\end{eqnarray}
      One can see that this Lagrangian contains a photon mass term 
      $\frac{1}{2} M_A^2 A_\mu A^\mu$ with $M_A= e v = - e \mu^2/\lambda$.
      There is a scalar particle $\eta$ 
      with a mass $M_{\eta}^2= 2 {v^2\lambda} = - 2\mu^2$,
      and there is a massless particle $\xi$ (a would-be Goldstone boson 
\cite{Djouadi:2005gi}) 
      which can be eliminated by the gauge transformation 
\cite{Higgs:1964pj,Higgs:1966ev,Kibble:1967sv}. 
      Indeed, there is a problem if one counts degrees of freedom
      in this theory. 
      At the beginning, one had 4 degrees of freedom,
      two for the complex scalar field $\phi$
      and two for the massless electromagnetic field $A_\mu$, 
      and now one has 5 degrees of freedom, one for $\eta$, 
      one for $\xi$ and three for the massive photon $A_\mu$. 
      Therefore, an unphysical field had appeared in the theory
      after the spontaneous violation of the local U(1) symmetry. 
      To find and to eliminate this field, one can notice the following.
      First, there is a ``suspicious'' bilinear term 
      $evA^\mu \partial_\mu \xi~$ in Lagrangian 
(\ref{Abelian-Lagrangian-1}), 
      which allows the vector field $A^\mu$       
      to directly transform to the scalar field $\xi$ during propagation.
      This means that $\xi$ plays a role of 
      the longitudinal component of the massive vector field $A^\mu$
      and one has to perform diagonalization to reach
      the physical eigenstate basis and to eliminate 
      this bilinear cross term
\cite{Kane:1987gb}.  
      Second, the diagonalization procedure in this particular
      case is exactly the gauge transformation 
(\ref{Ab-gauge}) which due to the U(1) gauge invariance
      eliminated completely the field $\xi$ from the Lagrangian.
      To illustrate the fact, one 
      can present the original complex scalar field $\phi$
      in the equivalent exponential form
      with the real $\eta'(x)$ and $\zeta(x)$
\begin{eqnarray}\label{Ab-phi-exp}
\phi(x)= \frac{1}{\sqrt{2}} (v +\eta +i \xi)  
    \equiv \frac{1}{\sqrt{2}} [v +\eta' (x) ] e^{i \zeta(x)/v}
\end{eqnarray}
      (in the first order 
      $\zeta=\xi$, $\eta=\eta'$ 
      due to $(v +\eta')(1+ i\zeta/v) = v + \eta' + i\zeta$)
      and use the freedom of gauge transformations 
      choosing exactly $\alpha(x) = -\zeta(x)/v$ in 
(\ref{Ab-gauge}).
      Therefore, (unitary) gauge transformation 
\begin{eqnarray} \label{Ab-unitary-gauge}
      \phi(x) \to e^{-i\zeta(x)/v} \phi(x), \qquad 
       A_\mu \to A_\mu - \frac{1}{ev} \partial_\mu \zeta(x)
\end{eqnarray}
       completely ``ate'' the phase factor $e^{i \zeta(x)/v}$ from 
(\ref{Ab-phi-exp}) and the scalar field has the simple form  
$$
\phi(x)= \frac{1}{\sqrt{2}} [v +\eta (x) ]. 
$$
       In the unitary gauge  
(\ref{Ab-unitary-gauge}) Lagrangian
(\ref{local-U1}) or
(\ref{Abelian-Lagrangian-1}) obtains the form (see Appendix
\ref{App-proves}):
\begin{eqnarray} \label{Ab-Lag-final}  
{\cal L} &=& 
       (\partial_\mu +ie A_\mu)\phi^*(\partial^\mu -ie A^\mu)\phi 
       - \mu^2 \phi^* \phi-\lambda(\phi^* \phi)^2
       -\frac{1}{4}F_{\mu \nu}F^{\mu \nu} =
\\ \nonumber 
&=&  \frac{1}{2}
      \partial_\mu\eta\partial^\mu\eta 
     -\frac{2v^2\lambda}{2}\eta^2 
     -\frac{F_{\mu\nu}F^{\mu\nu}}{4}
     +\frac{e^2v^2}{2}A_\mu A^\mu 
     +\frac{e^2}{2}A_\mu A^\mu (2 v \eta + \eta^2) 
     +\frac{v^4\lambda}{4}
     -{\lambda}v\eta^3 
     -\frac{\lambda}{4}\eta^4.
\end{eqnarray}
       There are no unphysical states in this Lagrangian at all.
       Furthermore, although Lagrangian
 (\ref{Ab-Lag-final}) has now the massive vector boson $A_\mu$,
       it is still gauge invariant, 
       because the initial Lagrangian
(\ref{local-U1}) was gauge invariant and only pure algebraic 
      transformation was carried out
\cite{Kane:1987gb}.

       This choice of gauge
       is called the unitary gauge.  
       The photon (with two degrees of freedom) has absorbed 
       the would-be Goldstone boson (with one degree of freedom) 
       and became massive (i.e., with three degrees of freedom), 
       the longitudinal polarization is the Goldstone boson. 
       The U(1) gauge symmetry is no longer apparent and we say that 
       it is spontaneously broken.  
       This is the Higgs mechanism 
\cite{Higgs:1964pj,Higgs:1966ev,Kibble:1967sv}
       which allows to generate masses for the gauge bosons:
      ``Gauge transformation ate the Goldstone boson''.
       The Higgs mechanism is clear from a 
       mathematical point of view, but its 
       physical interpretation is not yet completed in 
       the modern particle physics theory.
       One can see that a longitudinal state of the vector gauge boson, 
       which should exist for the massive boson 
       in the Lorentz-invariant theory (when one can 
       boost to the boson rest system),
       is the Goldstone boson which would exist if 
       the theory were not gauge invariant
\cite{Kane:1987gb}.

\subsection{The Higgs mechanism in the Standard Model}
\label{HM-SM}
      The Standard Model 
(SM) Lagrangian before EWSB has the form 
(see for example, 
(\ref{SM-Lagrangian}) and Appendix
\ref{App-SM}):
\begin{eqnarray}
{\cal L}_{\rm SM}= -\frac{1}{4} W_{\mu \nu}^a W^{\mu \nu}_a -\frac{1}{4} 
B_{\mu \nu}B^{\mu \nu} + \overline{L}\, i D_\mu \gamma^\mu \, L + 
\overline{e}_R \, i D_\mu \gamma^\mu \, e_R \ \cdots  
\end{eqnarray}
      For simplicity, the strong interaction part of it was here ignored. 
      In the non-Abelian SU(2)$\times$U(1) case of the SM 
      one needs to generate masses for the three gauge  bosons 
      $W^\pm$ and $Z$ but the photon should remain massless. 
      Therefore, one needs at least 3 degrees of freedom for 
      the scalar fields. 
      One would expect that the simplest choice is to use 
      an isovector state with exactly 3 scalar fields, but 
      in this case one lacks for massless fields and it is  
      impossible to generate all above-mentioned masses in the SM.
      In fact, one needs 
      a complex SU(2) doublet of scalar fields $\phi$ 
\begin{eqnarray}\label{Phi-def}
\Phi = \left(\begin{array}{c}\phi^+ \\ \phi^0 \end{array}\right) 
     = \frac{1}{\sqrt{2}} 
\left(\begin{array}{c} \phi_1 + i\phi_2 \\ \
                       \phi_3 + i\phi_4 \end{array} \right) 
\end{eqnarray}
      where $\phi_i$ are 4 real scalar fields (4 degrees of freedom).  
      The relevant scalar Lagrangian has the form
\begin{eqnarray}\label{SU2-ScalarL}
{\cal L} = (D^\mu \Phi)^\dagger (D_\mu \Phi)  - V(\Phi), 
\quad {\rm with} \quad 
V(\Phi) = \mu^2\Phi^\dagger\Phi + \lambda(\Phi^\dagger\Phi)^2,
\end{eqnarray}
      where both the product 
\begin{equation} \label{Phi-Phi}
\Phi^\dagger \Phi = 
          (\phi^{+*} \phi^{0*})
          \left(\begin{array}{c}\phi^+ \\ \phi^0 \end{array}\right) 
	 = \phi^{+*}\phi^{+} +
	   \phi^{0*}\phi^{0} =
	   \frac{1}{2} (\phi_1^2+ 
	   \phi_2^2 +\phi_3^2+\phi_4^2) = \frac{1}{2} \phi_i \phi^i
\end{equation}
      and, consequently, the potential $V(\Phi)$ 
      are invariant under the local gauge transformations
\begin{equation} \label{Phi-gauge}
\Phi(x) \to \Phi(x)' = e^{i\alpha_i(x)\tau_i/2}\Phi(x), 
\end{equation}
        where $\tau_i$ are Pauli matrices (see Appendix 
\ref{App-SM}) 
	and $\alpha_i(x)$ are transformation parameters.

        For $\mu^2 <0$, the potential $V(\Phi)$ has a minimum at 
$$
\Phi^\dagger \Phi = - \frac{\mu^2}{2\lambda} = \frac{v^2}{2}
$$
        and from 
(\ref{Phi-Phi}) one can conclude that there is an infinite number of 
       possible solutions of this equation.
       To preserve electric charge conservation (U$(1)_{\rm QED}$ symmetry), 
       this nonzero vacuum expectation value  
       should not be reached in the charged direction.
       A convenient choice of the neutral direction is
       $\phi_1= \phi_2= \phi_4= 0$ (see (\ref{Phi-def})).
       Therefore, the neutral component ($\phi_3$) 
       of the doublet field $\Phi$
      develops a nonzero vacuum expectation value 
\begin{eqnarray}
\langle\,\Phi\,\rangle_0 \equiv \langle 0|\Phi|0\rangle 
  = \frac{1}{\sqrt{2}}  \left( \begin{array}{c} 0\\ v 
  \end{array} \right) 
\quad {\rm with} \quad 
v= \left(-\frac{\mu^2}{\lambda} \right)^{1/2} .
\end{eqnarray}

        Now, as previously, using the pattern of the gauge symmetry of
(\ref{Phi-gauge}) one can write the field $\Phi$ in 
        the exponential form via four fields $\theta_{1,2,3}(x)$ and 
	$h(x)$: 
\begin{eqnarray}
\Phi(x) = 
       \frac{1}{\sqrt{2}}e^{i{\theta_a}(x)\tau^a(x)/v}
       \left(\begin{array}{c} 0 \\ (v + h(x))\end{array}\right). 
\end{eqnarray}
       Moving to the unitary gauge by means of a  
       proper gauge transformation of the field in the form
\begin{eqnarray} \label{Phi-Unitary}
\Phi(x) \to \Phi(x)' = e^{- i \theta_a (x) \tau^a (x)/v } \Phi(x) 
= \frac{1}{\sqrt{2}}\left(\begin{array}{c}0 \\ v+h(x)\end{array} \right) 
\end{eqnarray}  
       one ``gauges away'' three $\theta_a$ fields,
       chooses only one direction,    
       violates three global initial symmetries of the Lagrangian, 
       and leaves only one invariant 
(\ref{Phi-Phi}).
      For simplicity, in what follows for the field $\Phi(x)'$
      in the unitary gauge 
(\ref{Phi-Unitary})       the same notation $\Phi(x)$ will be used.  

      With $\Phi(x)$ from 
(\ref{Phi-Unitary}) one can expand (see Appendix 
\ref{App-proves}) the kinetic term 
      $(D^\mu \Phi)^\dagger (D_\mu \Phi) \equiv |D_\mu \Phi|^2 $ 
      of  Lagrangian 
(\ref{SU2-ScalarL}) 
\begin{eqnarray}\label{SM-Dmu}
|D_\mu\Phi|^2 &=& \left| 
                \left(\partial_\mu - ig_2 \frac{\tau_a}{2} W_\mu^a 
                 - i g_1 \frac{Y_H}{2}B_\mu \right)\Phi \right|^2 =
\\ 
&=& \nonumber
      \frac{1}{2}(\partial^\mu h)^2 
     +\frac{g^2_2}{8}(v+h)^2(W^\mu_1+i W^\mu_2)(W_\mu^1-i W_\mu^2)
     +\frac{1}{8}(v+h)^2\left({g_2}W^\mu_3-{g_1Y_H}B^\mu\right)^2 =
\\&=& \label{SM-Dmu-mass}
      \frac{g^2_2v^2}{8}(W^\mu_1+i W^\mu_2)(W_\mu^1-i W_\mu^2)
     +\frac{v^2}{8}\left({g_2}W^\mu_3-{g_1Y_H}B^\mu\right)^2 
     +\frac{1}{2}(\partial^\mu h)^2 + ... 
\end{eqnarray}
     The first term in 
(\ref{SM-Dmu-mass}) 
     is the mass term $M_W^2 W^+_\mu W^{-\mu}$ 
     for the charged gauge boson field 
\begin{equation}\label{W-mass}
W^\pm = \frac{1}{\sqrt{2}} (W^1_\mu \mp i W^2_\mu)
\quad {\rm with} \quad
M_W =\frac{1}{2}vg_2. 
\end{equation}
      In particular, the last relation allows one to fix 
      the vacuum expectation value $v$ in terms of the $W$ 
      boson mass $M_W$ and the Fermi constant $G_{\rm F}$ 
      (Appendix \ref{App-SM})
\begin{eqnarray}
M_W=\frac{g_2v}{2}=
\left(\frac{\sqrt{2}g_2^2}{8 G_{\rm F}} \right)^{1/2} \qquad 
{\rm and }\qquad
v= \frac{1}{(\sqrt{2} G_{\rm F})^{1/2} } \simeq 246~{\rm GeV}.
\label{MW-vs-G-v}
\end{eqnarray}

     The second term in 
(\ref{SM-Dmu-mass}) mixes two neutral components of the gauge fields 
     $W^\mu_3$ and $B^\mu$, but after diagonalization 
     (moving to mass eigenstates $Z_\mu$ and $A_\mu$) in the form 
\begin{eqnarray}\label{ZA-fields}
Z_\mu = \frac{g_2 W^3_\mu- g_1 B_\mu}{\sqrt{g_2^2+g_1^2}},
\qquad
A_\mu = \frac{g_2 W^3_\mu+ g_1 B_\mu}{\sqrt{g_2^2+g_1^2}}  
\end{eqnarray}
      one can interpret it as  a mass term
      $\frac{1}{2} M_Z^2 Z_\mu Z^\mu$ with 
\begin{equation}\label{Z-mass}
M_Z= \frac{1}{2} v  \sqrt{g^2_2+g_1^2}.
\end{equation}
      Here $Y_H=1$ was used.
      It is very important that the neutral field 
      $A_\mu$, being orthogonal to $Z_\mu$, has no mass term at all.
      The term like $\frac{1}{2} M_A^2 A_\mu A^\mu$ does not appear.

     Therefore, 
     by spontaneously breaking of the symmetry 
     ${\rm SU(2)_L\times U(1)_Y \to U(1)_{Q}}$ 
     (from 4 generators to only 1), three 
     Goldstone bosons have been absorbed by the $W^\pm$ and $Z$ bosons 
     to form their longitudinal components and get their masses. 
     Since the ${\rm U(1)_{Q}}$ symmetry is still unbroken, 
     the photon which is its generator remains massless.

In fact the photon (the gauge boson of the ${\rm U(1)_{Q}}$ symmetry) 
remains massless and the symmetry is still unbroken due to the fact that 
the Lagrangian and the vacuum field 
$\Phi_0=\langle\,\Phi\,\rangle_0$
of the system both and 
simultaneously remain invariant under a U(1) transformation, 
which is a direct consequence 
of the electric charge conservation
(which is observable and must be hold in any system 
after any correct transformations).
Indeed, the electric charge of the Higgs field $Q$ is connected 
with the eigenvalue of the weak SU(2) isospin operator $T_3\equiv \tau_3$
and U(1) hypercharge for the Higgs field $Y_H$ by means of the simple
relation (Appendix \ref{App-SM})
\begin{equation}\label{Q-T-Y}
Q = T_3 + \frac{Y_H}{2}.
\end{equation}
      Since we have already fixed the charge of the lower SU(2) component of
      $\Phi$ (vacuum is neutral) and for this component 
      $T_3=-\frac12$, we conclude that $Y_H=1$. Applying relation (\ref{Q-T-Y})
      to the upper ($T_3=\frac12$) component of the doublet $\Phi$, one deduces
      that it is positively charged (this justifies our notation 
      in (\ref{Phi-def})).

       It is interesting to notice that
       the vacuum {\em is charged} under the initial 
       SU(2) and U(1), and violates these symmetries. 
       ``Fortunately'', 
       the vacuum has zero eigenvalue of the electric charge operator
       $Q\Phi_0 = (T_3 + \frac{Y_H}{2})\Phi_0 = 0$ 
       and is, therefore, invariant under 
       the ${\rm U(1)_{Q}}$ symmetry transformation
$$
\Phi_0 \to \Phi^{'}_0 = e^{i\beta(x)Q}\Phi_0 = \Phi_0.
$$
      
\subsection*{Fermion mass generation}
       The arrangement of  scalar Higgs fields $\phi$ 
       in the complex SU(2) doublet 
(\ref{Phi-def}) 
       allows one to construct ${\rm SU(2)_L \times U(1)_Y}$  
       invariant interaction
       of the Higgs fields with fermions, being
       only the SU(2) doublets or singlets.
       For leptons and down-type quarks of all generations
       this ${\rm SU(2)_L \times U(1)_Y}$ 
       invariant Yukawa Lagrangian has the form
\begin{eqnarray}\label{SM-Yukawa-d}
{\cal L}_d =-\lambda_{e} 
          \left(\bar{L}\,\Phi\,e_{R}+\Phi^\dagger\,\bar{e}_{R}\,L\right) 
          -\lambda_{d}
          \left(\bar{Q}\,\Phi\,d_{R}+\Phi^\dagger\,\bar{d}_{R}\,Q\right). 
\end{eqnarray}
           The second terms in each bracket are 
	   relevant Hermitian conjugates.
	   It is important to note that 
	   with the field $\Phi$ 
	   which has $Y_H=1$, 
           the total hypercharge of each term in 
(\ref{SM-Yukawa-d}) equals zero due to 
         $Y_{L_i}=-1$, $Y_{e_{R_i}}=-2$, $Y_{Q_i}=\frac{1}{3}$, 
	 and $Y_{d_{R_i}}= -\frac{2}{3}$
(see eq.~(\ref{Hypercharges}) in 
Appendix~\ref{App-SM}). 
         On the contrary,  
	 if one uses the Yukawa term in the form 
	 $~\bar{Q}\,\Phi\,u_{R}~$
	 with the same 
	 $\Phi$ field 
	 ($Y_H=1$) for up-type quarks,
         one arrives at hypercharge violating Lagrangian  
	 due to the fact that 
	 $Y_{u_{R_i}}= \frac{4}{3}$ and
	 $-\frac{1}{3} + 1 +\frac{4}{3}=2\ne 0$.
	 To bypass the problem, one should use the isodoublet  
	 $\tilde{\Phi}= i\tau_2 \Phi^* = 
\left(\begin{array}{c}\phi^{0*} \\ -\phi^{-}\end{array}\right)
\to \frac{1}{\sqrt{2}}\left(\begin{array}{c}v+h(x)\\ 0\end{array}\right)$ 
	 which has hypercharge $Y=-1$ due to complex conjugation. 
	 For up-type quark SM 
	 Yukawa interaction is 
\begin{eqnarray}\label{SM-Yukawa-u}
{\cal L}_u = - \lambda_u 
          \left(\bar{Q}\,\tilde{\Phi}\,u_R 
          +\tilde{\Phi}^\dagger\,\bar{u}_{R}\,Q\right). 
\end{eqnarray}
     Therefore, after the EWSB when the Higgs field 
(\ref{Phi-def}) 
     has obtained the nonzero vev 
     one can generate masses for all fermions of the SM
     via the interaction Lagrangians
(\ref{SM-Yukawa-d}) and 
(\ref{SM-Yukawa-u}).

     Consider, for instance, the case of the electron (the first term in 
(\ref{SM-Yukawa-d})).
     With the Higgs field in the unitary gauge
(\ref{Phi-Unitary}), one obtains
\begin{eqnarray}\nonumber
{\cal L}_e 
    &=&-\lambda_{e}
    \left(\bar{L}\,\Phi\,e_{R}+\Phi^\dagger\,\bar{e}_{R}\,L\right) =
   -\frac{\lambda_{e}}{\sqrt{2}}\, 
     (\bar{\nu}_e,\bar{e}_L)\, 
     \left(\!\begin{array}{c}0\\v+h\end{array}\!\right)\!\,e_{R}
   -\frac{\lambda_{e}}{\sqrt{2}}\, (0, v+h )\,      \bar{e}_{R}\,
     \left(\!\begin{array}{c} {\nu}_e
     \\ {e}_L  \end{array}\!\right) = 
\\ \label{Yukawa-eH}
&=&   -\frac{\lambda_e\,v}{\sqrt{2}}\,
        (\bar{e}^{}_L\,e^{}_R+\bar{e}^{}_R\,e^{}_L) 
      -\frac{\lambda_e}{\sqrt{2}}\,
        (\bar{e}^{}_L\,e^{}_R+\bar{e}^{}_R\,e^{}_L)\, h .
\end{eqnarray}
      Taking into account that 
      $\bar{\psi}^{}_L\,\psi^{}_R+\bar{\psi}^{}_R\,\psi^{}_L
      = \bar{\psi}\,\psi~$ 
(see (\ref{Intro-m_e}))
      one can conclude that the first term in 
(\ref{Yukawa-eH}) looks exactly as a mass term for fermions 
      $-m\bar{\psi}\,\psi$, 
      with the electron mass 
      (and in complete analogy for the up- and down-quarks)
\begin{eqnarray}\label{Yukawa-masses}
m_e= \frac{\lambda_e\, v}{\sqrt{2}}, \quad 
m_u= \frac{\lambda_u\, v}{\sqrt{2} }, \quad 
m_d= \frac{\lambda_d\, v}{\sqrt{2}}. 
\end{eqnarray}
       Due to unknown values of the Yukawa constants 
       $\lambda_{e,u,d}~$ it is impossible to calculate
       the masses of electron and quarks, but if one knows 
       these masses from experiment, it is 
       possible to estimate the strength of the
       electron-electron-Higgs 
       (and any fermion-fermion-Higgs) 
       interaction (see the second term in 
(\ref{Yukawa-eH})) inverting formulae
(\ref{Yukawa-masses}):
\begin{equation}\label{ffh}
{\cal L}_{eeh} =  
      -\frac{m_e}{v}\,\bar{e}\,e\,h 
      -\frac{m_u}{v}\,\bar{u}\,u\,h 
      -\frac{m_d}{v}\,\bar{d}\,d\,h + ... 
\end{equation}
      A very important consequence of the 
      fermion-fermion-Higgs interaction
(\ref{ffh})
      is its direct dependence of the fermion mass.
      The larger the mass the stronger 
      this interaction.

\subsection*{The Higgs boson}
    The kinetic part of the Higgs field, $\frac{1}{2} (\partial_\mu h)^2$, 
    comes from the covariant derivative $|D_\mu \Phi|^2$
    (the last term in 
(\ref{SM-Dmu-mass})), while the Higgs mass and Higgs self-interaction
    parts, come (as it should be) from the scalar potential 
    $V(\Phi)=\mu^2\Phi^\dagger\Phi+\lambda (\Phi^\dagger \Phi)^2$
(\ref{SU2-ScalarL}) which after EWSB 
    takes the form
\begin{eqnarray*}
V(h) &=&\frac{\mu^2}{2} (0,\, v+h) 
      \left(\!\begin{array}{c}0\\v+h\!\end{array}\right) 
      +\frac{\lambda}{4}
      \left|(0,\, v+h)
      \left(\!\begin{array}{c}0\\v+h\!\end{array}\right)\right|^2 =
\frac{\mu^2}{2}
       (v+h)^2 + \frac{\lambda}{4}(v+h)^4, 
\end{eqnarray*}
       Finally, with the relation $\mu^2=-v^2\lambda$
       the pure SM Higgs Lagrangian is given by 
\begin{equation}\label{SM-pure-H}
{\cal L}_{h} 
   =\frac{1}{2}(\partial^\mu h)^2 
   -\frac{2\lambda v^2}{2}\, h^2 - \lambda v \, h^3 
   -\frac{\lambda}{4}\, h^4  
   +\frac{\lambda v^4}{4}.
\end{equation}
      This Lagrangian coincides with the simple scalar Lagrangian
(\ref{Lagr-scalar-sigma}) 
      and despite the presence of the cubic term $\lambda v h^3$ 
      it has vacuum state at $h(x)=0$ (see Section 
\ref{simplest}).
      From Lagrangian 
(\ref{SM-pure-H}) one can conclude that the Higgs boson mass is 
\begin{equation}\label{SM-H-mass}
      M_h^2=2 \lambda v^2. 
\end{equation} 
      The strength of the Higgs self-interactions
      is proportional to the square of the Higgs mass
$$ 
g_{h^3}= \lambda v = \frac{M_h^2}{2 v}, \qquad 
g_{h^4}= \frac{\lambda}{4} = \frac{M_h^2}{8 v^2}.
$$
      In accordance with relation 
(\ref{ffh}), the interaction of the Higgs boson with a fermion
      is proportional to the mass of this fermion
$\displaystyle g_{hff}=\frac{m_f}{v}$. 
      Furthermore, 
      the Higgs boson couplings to the gauge bosons come from relation  
(\ref{SM-Dmu}) in almost full analogy with the vector boson mass terms
\begin{eqnarray}\nonumber
{\cal L}_{hVV}
&=&     M_W^2\left(1+\frac{h}{v}\right)^2 W^+_\mu W^{-\mu}
     +\frac{M_Z^2}{2}\left(1+\frac{h}{v}\right)^2  Z_\mu Z^\mu.
\end{eqnarray}
      Here the gauge boson mass definitions 
(\ref{W-mass}) and (\ref{Z-mass}) were used.
      Thus, again 
      the Higgs boson couplings to the gauge bosons are
      proportional to the squared mass of these bosons 
\begin{eqnarray}\label{SM-g_hhWW}
g_{hWW}=2\frac{M_W^2}{v}, \quad g_{hhWW}=\frac{M_W^2}{v^2}, 
\quad {\rm and} \quad 
g_{hZZ}=\frac{M_Z^2}{v}, \quad g_{hhZZ}=\frac{M_Z^2}{2v^2}.
\end{eqnarray}

\smallskip
      Therefore, {\em the only one}\/ 
      isodoublet $\Phi$ of scalar fields 
      allows mass generation for all massive 
      particles of the Standard Model --- 
      the weak vector bosons $W^\pm$,  $Z$, fermions, and 
      the Higgs boson itself, while 
      preserving the SU(2)$\times$U(1) gauge symmetry in the 
      spontaneously broken or hidden form. 
       The electromagnetic ${\rm U(1)_{Q}}$ symmetry, 
       due to the requirement of electric charge conservation, and
       the SU(3) color symmetry, due to color charge conservation,  
       both remain unbroken. 

\smallskip
    Nevertheless, despite this beautiful picture, 
    the problem of the Higgs boson mass still remains unclear. 
    Indeed,  
    the mass of the Higgs boson is generated by 
    the Higgs self-interaction and is 
    defined by the parameter $\lambda$, 
    the coupling of Higgs self-interaction. 
    There is no clear idea within the SM 
    concerning the source of  $\lambda$, 
    and its value stays, in principle, undefined
    (together with the Higgs boson mass $M_h$).
    What makes the situation much worse is that
    today there is no any other observable 
    which could depend on  $\lambda$ and
    which could give a way to measure it experimentally
\cite{Kane:1987gb}.

    In the next section, one can find some 
    review of the available information about  
    this ``mysterious'' $\lambda$ parameter. 

\section{On Higgs mass and self-interaction}
\subsection{\boldmath The case $\lambda=\lambda_t=1$}
    It is not necessary to claim that 
    the Higgs boson and the top quark are the key 
    ingredients of the SM. 
    It is also well known that the SM 
    cannot predict their masses directly.
    Therefore, any idea about values of the top quark 
    and the Higgs boson masses has the right to some attention.
 
    On this way, consider first a very simple case based
    on 
    the assumption that both the Higgs self-coupling $\lambda$
    and the 
    Yukawa top coupling $\lambda_t$ are equal to 1 at the electroweak scale.
    This assumption 
{\em (proposed by N. Giokaris)}\/ 
    surprisingly allows one to obtain rather 
    accurate predictions for the top quark and Higgs boson masses
    $m_t$ and $M_h$.

    With the Fermi constant value 
$G_{\rm F} = (1.16637 \pm 0.00001) \cdot 10^{-5}~{\rm GeV}^{-2}$
(see eq.~(\ref{GFermi}) in Appendix \ref{App-SM})
    one obtains  
    for Higgs vacuum expectation value $v$ (see eqs. 
(\ref{MW-vs-G-v}) or
(\ref{MW-vs-v})) 
\begin{equation}\label{NG-v}
     v = \sqrt{\frac{1}{\sqrt{2}G_{\rm F}}} = 246.221~{\rm GeV}.
\end{equation}
     As it follows from 
(\ref{Yukawa-masses}),
     the mass of a fermion $f$ is defined by the fermion-Higgs 
     Yukawa coupling $\lambda_f$. 
     Therefore, 
     if one assumes that the maximal value of $\lambda_f$ is 
     equal to 1,
     then the heaviest possible fermion mass 
     appeares just right equal to the 
     top quark mass ($\lambda_t=1$) 
$$
m_t = \frac{\lambda_t\, v}{\sqrt{2}} =
m^{\max}_f  = \frac{v}{\sqrt{2}} = 174.105~{\rm GeV}.
$$
     This value of the top quark mass coincides (within errors)
     with $m_t = 172.7\pm 2.8~$GeV, which was used in the 
     fit of all precision data by PDG-2006
\cite{Yao:2006px} including all involved radiative corrections.
     In particular, the result of this fit was the 90\%
     confidence level for the Higgs mass
\begin{equation}\label{Exp-M_h}
46~{\rm GeV} \le  M_h \le  154~{\rm GeV}. 
\end{equation}

     There is, however, another possibility 
     for determination of $M_h$.
     Having in mind that the Higgs scalar field, 
     through the SM Higgs mechanism, gives the SM
     particles their masses, 
     it would be natural to assume that 
     the Higgs particle should be heavy enough 
     to have a chance (at least in principle)
     to decay also into a real $\bar{t}t$-pair.
     Therefore, the Higgs boson mass should not be smaller than
\begin{equation}\label{NG-mH}
    M_h=2 m_t 
    = 2\times 174.1~{\rm GeV} = 348.2~{\rm GeV}.
\end{equation}
    One can see that this value can be obtained directly from 
    the Higgs mass definition
(\ref{SM-H-mass})
$ M_h = \sqrt{2 \lambda v^2} $\/ 
    by taking the Higgs self-coupling $\lambda=1$.

    Therefore, 
    assuming that the Yukawa coupling of the heaviest particle 
    to the Higgs field is equal to 1, one can obtain 
    $m_t= 174.105~{\rm GeV}$
    in very good agreement
    with the latest Tevatron result. 
    Assuming that the self-coupling in the Higgs
    potential is equal to 1, one can obtain 
    $M_h = 348.2~{\rm GeV}$
    or $M_h = 2m_t$.  
   Finally, 
\begin{equation}\label{NG-final}
m_t = \frac{v}{\sqrt{2}}, \quad  
M_h = 2 m_t = \sqrt{2} v, \quad
v^2 = m_t \times  M_h = 2 m^2_t = \frac{M^2_h}{2}.
\end{equation}
      Note that now 
      the vacuum expectation value is the geometric mean of
      the top-quark and the Higgs particle masses.
      Relations 
(\ref{NG-final}) tie together the two assumptions we made about
      the top quark Yukawa couplings and the $\lambda$ 
      parameter of the Higgs field.
      In principle, one can look for 
      any deeper symmetry or 
      other arguments trying to justify these assumptions.

\subsection{\boldmath Constraints from triviality} 
     With vacuum value $v$ from 
(\ref{MW-vs-G-v}) one can write the Higgs mass 
(\ref{SM-H-mass}) in the form
\cite{Aitchison:2003ji}
\begin{equation}\label{Tr-A-M_h}
M_h = v \sqrt{2\lambda} = 
2 \sqrt{\lambda} \times 174.105~{\rm GeV}.
\end{equation}
     If a dimensionless constant $\lambda$ is $O(\alpha)$ 
     one has a perturbative theory, 
     while if it is $O(1)$, one would say the theory is strongly coupled.
     From 
(\ref{Tr-A-M_h}) and the present experimental 
     bounds on $M_h$
(see, for example (\ref{Exp-M_h})), 
     one concludes 
     that we are already not {\em very}\/ 
     far from the strongly coupled region
\cite{Aitchison:2003ji} and 
     the following question is 
     reasonable:
     Can $\lambda$ (the renormalized coupling) take {\em any}\/ 
     value  at all? 
     That is, can $M_h$ (for fixed $v$) be arbitrarily large?
       
     To answer this question one has, first of all,  
     to recall that in a 
     renormalizable theory the value of $\lambda$ (as well as
     the value of $\mu^2$) has to be defined
     at a certain scale and the value at another scale is 
     different (i.e., $\lambda$ `runs'). 
     At the one-loop level, the renormalization group equation for
     the Higgs 
	quartic self-coupling $\lambda$ is given by  
(see, for example,      
\cite{Grojean:2005ud}):
\begin{equation}\label{Tr-RG}
16\pi^2 \frac{\lambda}{d \ln{E}}
\equiv \beta_\lambda(\lambda)
=
24\lambda^2 - (3g_1^2+9g^2_2 -12 \lambda^2_t)\lambda 
              + \frac{3}{8}g_1^4 
              + \frac{3}{4}g^2_1 g^2_2 
	      + \frac{9}{8}g^4_2 
	      -6\lambda^4_t + ....
\end{equation}
     For a rather large $\lambda$ the first term dominates and
     forces $\lambda$ (together with the Higgs mass) to increase
     infinitely with energy scale $E$. 
     In this regime, the solution of 
(\ref{Tr-RG}) is 
\begin{equation}\label{Tr-L-Pole}
\lambda(E) = \frac{\lambda} 
            {1-\frac{3}{2\pi^2}\,\lambda\, 
	    \ln\left({E}/{v}\right)}.
\end{equation}
     The  ``physical'' $\lambda$ is defined at the scale $E=v$. 
     It follows from 
(\ref{Tr-L-Pole}) that the theory breaks 
     down --- exhibits the so-called Landau pole ---
     at the energy scale  
\begin{equation}\label{Tr-E-Pole}
E^* \simeq v \exp{\left(\frac{2\pi^2}{3\lambda} \right)}
     = v \exp{\left(\frac{4\pi^2 v^2}
               {3 M^2_h} \right)}.
\end{equation}
     More conservatively one can say that $\lambda(E)$ becomes so 
     large that all perturbative expectations are meaningless. 
      Here the Higgs mass $M_h$ definition (\ref{Tr-A-M_h})
      was used. 
      In fact, relation 
(\ref{Tr-E-Pole}) 
      gives the upper bound for the cut-off scale of the SM
$~\Lambda\le E^*~$. 
      Above the scale 
$\Lambda$,  
      some new physics should appear to prevent
      this ``blowup''. 
      Formula 
(\ref{Tr-E-Pole}) is very remarkable, because it exhibits  
      exponential sensitivity to the unknown $M_h$.
      For rather small Higgs masses the breakdown scale is high --- 
      for $M_h \simeq $~150~GeV $E^* \simeq 6\times 10^{17}$~GeV. 
      However, for $M_h\simeq$~700 GeV, $E^* $ is already as low as 1~TeV. 
      Clearly, at such a value of $M_h$, 
      the Higgs mass is essentially equal to the
      `breakdown scale' itself 
      and $M_h$ cannot get any higher without new physics
      (some non-perturbative phenomena, or, perhaps, supersymmetry)
\cite{Aitchison:2003ji}.
       Therefore, for a fixed value of the SM cut-off  
$\Lambda = E^*$ relation 
(\ref{Tr-E-Pole}) gives an upper bound on the Higgs mass.
       In particular, one cannot take $\lambda (E)\to \infty$, since
       in this case one necessarily has $\lambda=0$ 
       and, therefore, no any EWSB can occur
\cite{Grojean:2005ud}.

      The last observation is realization of the general 
      ``Triviality problem''.  
      It was theoretically shown  
(see, for example
\cite{Callaway:1983zd,Callaway:1988ya}) 
       that a pure $\phi^4$ scalar field theory with Lagrangian 
(\ref{Lagr-scalar}) 
\begin{equation}\label{TR-Lagr-scalar}
{\cal L} = \frac12 (\partial_\mu \phi)^2 
         - \frac{\mu^2_0}{2}\phi^2
          -\frac{\lambda_0}{4}\phi^4 
\end{equation}
       is trivial in four space-time dimensions. 
       The word ``trivial'' here means 
       that the scalar field does not interact with itself. 
       Triviality is equivalent to the statement that the renormalized
       quartic coupling        
$\lambda_r = \lambda(E)$ 
(\ref{Tr-L-Pole}) is equal to zero. 
       In other words, the scalar particles interact in such a 
       (strong enough) way 
       as to screen totally any bare charge $\lambda_0$, or 
       given the low-energy value of
       the Higgs coupling, the Higgs coupling will eventually blow up
       at some finite momentum scale $\Lambda_L$ (the Landau pole). 
       The stronger is the low-energy Higgs coupling,  the smaller is 
       $\Lambda_L$.

       This triviality seemingly persists for {\em all}\/ 
       values of the bare coupling constant 
       and, therefore, presumably precludes the existence 
       of spontaneous symmetry breaking in the pure $\phi^4$ field theory
\cite{Callaway:1988ya}.
       At first glance, this claim looks very surprising, 
       but it is a direct result of the renormalization group equation 
(\ref{Tr-RG}) 
       for the effective (or "running") quartic constant $\lambda(E)$. 
       Equation
(\ref{Tr-RG}) has the following boundary conditions:
\begin{equation}\label{TR-Bounds}
\lambda(E=v)=\lambda_r, \qquad {\rm and} \qquad
\lambda(E\to\infty) = \lambda_0, 
\end{equation}
    where $\lambda_r$ is the renormalized quartic coupling constant
    at the electroweak scale. 
    Conditions
(\ref{TR-Bounds}) 
    simply state that at high-momentum transfer (or energy) an incident
    (scalar) particle interacts with the bare charge of the target 
    (scalar) particle.

    In the language of the renormalization group, the triviality of a 
    $\phi^4$ theory is essentially equivalent to two statements.
    First, the bare coupling constant $\lambda_0$ is finite. 
    Second, the beta function $\beta_\lambda(\lambda)$ 
     is positive and equals zero only when $\lambda$ is zero.
     These two statements 
     imply that the renormalized coupling 
     $\lambda_r$ is zero for any sensible (i.e. finite and positive) value 
     of the bare coupling $\lambda_0$.
     This is the evidence for the triviality 
     of $\phi^4$ theory
\cite{Callaway:1983zd,Callaway:1988ya}.
      Theory becomes always meaningless for $\lambda_0\neq0$
      (it has sense only if $\lambda_0=0$), 
      but this means total
      absence of any $\lambda\phi^4$ self-interactions. 

     One can see that 
     there are two bare parameters, $\mu_0$ and $\lambda_0$
     in the original classical Lagrangian 
(\ref{TR-Lagr-scalar}). 
      When quantum effects are accounted for (i.e., when the theory is 
      renormalized), all that remains is one parameter, 
      the renormalized mass $\mu_r$. 
      Quantum effects have determined that $\lambda_r$ is zero. 
      Therefore, in the SM one parameter, the Higgs mass, 
      is not determined by low-energy phenomenology
      in the classical (tree-level) approximation. 
      In fact, the renormalization effects may 
      generally bound the Higgs mass from above
\cite{Callaway:1983zd,Callaway:1988ya}.
      Indeed, due to this triviality of the $\phi^4$ theory the SM is
      inconsistent as a fundamental theory but is a reasonable
      effective theory with momentum cut-off $\Lambda$.
      Furthermore, by requiring that $\Lambda$ be larger than 
      the Higgs mass in order to maintain the consistency of the
      SM as an effective theory
(see comments to eqns.
(\ref{Tr-L-Pole}) and (\ref{Tr-E-Pole}))
       one can derive the so-called 
       triviality upper bound on the Higgs mass in the SM.
       This upper bound of 1~TeV was for the first time obtained in 
\cite{Dashen:1983ts}.
       These (triviality)
       arguments were also successfully used to obtain 
       the Higgs mass upper bounds in some SUSY extension for the SM
(see, for example 
\cite{Wu:1994eb,Choudhury:1995bx}).
	In particular, the absolute upper bound on the 
	lightest Higgs mass was obtained as $2.8 M_W$ 
	by requiring that the Higgs couplings remain finite at 
	beneath the momentum cut-off $\Lambda$ for the NMSSM
\cite{Wu:1994eb}.

        There is another possibility to solve 
	(against the triviality) the Higgs mechanism 
	with fundamental scalars --- 
	a new phenomenon must occur in the 
	theory when gauge fields are present.
	Following Callaway
\cite{Callaway:1983zd}, 
        consider the effect of coupling a gauge field to the scalar field of 
(\ref{TR-Lagr-scalar}). 
        It is demonstrated in 
\cite{Callaway:1983zd} that for the
        combined theory to be nontrivial, the 
	renormalized quartic coupling $\lambda_r$
	must not be too strong. 
	The breakdown of total screening 
	(entering in nontrivial regime) occurs 
	when the quartic coupling constant $\lambda_r$
	is less than the effective quartic coupling
	generated by the gauge field interaction 
\begin{equation}\label{TR-Gauge}
\lambda_r \le \xi\, g^2_r. 
\end{equation}
      Here $g_r$ denotes the renormalized gauge coupling constant
      and $\xi\ge 0$ is some calculable constant. 
       In the SM  the squared ratio of
       Higgs to W-boson mass is given at the tree level by
\begin{equation}\label{TR-WH}
\left(\frac{M_h}{m_W}\right)^2 = 
8 \frac{\lambda_r}{g^2_r} \le 12.8. 
\end{equation}
    for reasonable parameter choices, and similarly for other theories
\cite{Callaway:1983zd,Callaway:1988ya}.
        In particular, taking, for example, $g^2_r = g^2_2 = 0.446$, 
	from 
(\ref{TR-WH}) one has $\lambda_r \le 0.72$. 

        Therefore, 
the assumption that a scalar field theory without gauge fields 
is trivial (i.e., that the renormalized quartic coupling is zero) 
implies strong constraints on a theory with gauge fields. 
The addition of gauge fields can in fact make a trivial pure scalar
theory nontrivial. 
Indeed, such a phenomenon may occur in realistic theories such
as the standard model of the weak interaction and in grand unified theories.
The mechanism by which triviality is eliminated typically 
works for a small range of renormalized coupling constants of the theory. 
Basically, a bare scalar particle screens itself totally, 
so that the renormalized scalar charge is zero regardless its bare value. 
The addition of a gauge field generates an effective quartic coupling
constant.
If this effective coupling is at least as large as the original 
coupling it can destroy the total screening of the bare charges.
However, the screening persists if the quartic coupling is 
much larger than the effective (gauge + quartic) coupling. 
The necessity of the destruction of the screening
phenomenon forces restrictions on the bare couplings.
This restriction in turn implies a calculable upper
bound on the ratio of Higgs to gauge boson mass 
(\ref{TR-WH}).
For details see 
\cite{Djouadi:2005gi,Callaway:1983zd,Callaway:1988ya}.

\subsection{Constraints from unitarity}
       To obtain unitary bound on the 
       Higgs mass (and $\lambda$), one has to use the
       decomposition of the scattering amplitude into the
       partial waves 
\cite{Lee:1977eg,Chanowitz:1985hj,Djouadi:2005gi}:
\begin{equation}\label{TR-partial}
A = 16\pi \sum^{\infty}_{l=0} (2l+1) P_l(\cos\theta)\, a_l, 
\end{equation}
      where $P_l (\cos\theta)$ are the Legendre polynomials, and 
      the partial wave amplitudes $a_l$ of orbital angular momentum $l$
      are given by 
$$
a_l = \frac{1}{32\pi}\int^{1}_{-1}d(\cos\theta)P_l(\cos\theta)\,A.
$$
       The differential and total cross sections have the forms
\begin{equation}\label{TR-cross-sections}
\frac{d\sigma}{d\Omega} = \frac{1}{2\pi}\frac{d\sigma}{d\cos\theta}
= \frac{|A|^2}{64\pi^2\,s}
\qquad{\rm and}\qquad
\sigma = \frac{16}{s}\sum^{\infty}_{l=0}(2l+1)|a_l|^2.
\end{equation}
      Here the orthogonality property of the Legendre polynomials
      in the form 
$\int d \cos\theta P_l P_{l'}=\delta_{l l'}$ was used. 
        One knows that due to the optical theorem 
	the total cross section is proportional to the imaginary
	part of the amplitude $A$ in the forward direction 
	($\theta=0$); therefore, 
\begin{equation}\label{Optic}
\sigma = \frac{1}{s} \, {\rm Im}\left(A(\theta=0)\right)
=\frac{16\pi}{s} \sum_{l=0}^{\infty}(2l +1) |a_l|^2. 
\end{equation}
      With 
(\ref{TR-partial}) from 
(\ref{Optic}) one has that 
${\rm Im}(a_l)=|a_l|^2=|{\rm Re}(a_l)|^2+|{\rm Im}(a_l)|^2$ 
or 
$|{\rm Re}(a_l)|^2+|{\rm Im}(a_l)|-\frac{1}{2}|^2 = \frac{1}{4}.$
        This is the equation of a circle of radius $\frac{1}{2}$
	with the center at $(0, \frac{1}{2})$.
	Therefore, the real part lies between 
        $-\frac{1}{2}$ and $\frac{1}{2}$, 
	and one finally has
\cite{Grojean:2005ud,Djouadi:2005gi}
\begin{equation}\label{TR-optical}
|{\rm Re}(a_l)| \le \frac12. 
\end{equation}
      With the Higgs boson contribution to the scattering amplitude,
      which cancels a dangerous energy growth of the amplitude, 
      one gets 
\cite{Grojean:2005ud,Djouadi:2005gi}
\begin{equation}\label{TR-a0}
a_0 =\frac{g^2_2 M^2_h}{64\pi M^2_W}.
\end{equation}
       This leads to the upper bound for the Higgs mass $M_h \le 1.2~$TeV.
       In fact, with extra channels including only $W$ and $Z$
       gauge bosons one has a more stringent bound
\cite{Lee:1977eg,Chanowitz:1985hj}
$$M_h \le 780~{\rm GeV}
\qquad{\rm and}\qquad
\lambda \le 5. 
$$ 
      These bounds give an order of magnitude estimate and
      they should not be considered as tight bounds  
\cite{Grojean:2005ud}.

\subsection{Constraints from stability}
     It is clear that $\lambda$ could not be very small, 
     otherwise the typical Higgs potential 
(\ref{Lagr-scalar}) will not be constrained
     from below and the theory will loose its stability.
     At the low mass Higgs limit (low $\lambda$ limit) 
     in the renormalization group equation 
(\ref{Tr-RG})      
     the top Yukawa coupling 
     $\lambda_t$ dominates which forces  $\lambda$ 
     (and the Higgs boson mass) to decrease with energy increase:
\begin{equation}\label{ST-RG}
16\pi^2 \frac{\lambda}{d \ln{E}} =  -6\lambda^4_t + ...
\end{equation}
      To obtain the energy dependence of $\lambda$ 
      in this case, one needs a renormalization group equation 
      for the top Yukawa coupling. 
      At the one-loop approximation it can be given as 
\begin{equation}\label{ST-RG-top}
16\pi^2 \frac{\lambda_t}{d \ln{E}}=
	      \frac{9}{2}\lambda^3_t + ...
\end{equation}
       The solution of both the renormalization group equations
(\ref{ST-RG}) and (\ref{ST-RG-top}) is 
\cite{Grojean:2005ud}: 
\begin{eqnarray}\label{ST-RGE}
\lambda^2_t(E) = \frac{\lambda^2_0}
                  {1-\frac{9}{16\pi^2}\lambda^2_0\ln(E/E_0)}
\qquad {\rm and} \qquad 
\lambda(E) = \lambda_0 - 
           \frac{ \frac{3}{8\pi^2} \lambda^4_0\ln(E/E_0)}
                {1-\frac{9}{16\pi^2}\lambda^2_0\ln(E/E_0)}.
\end{eqnarray}
     For rather large $E$, the Higgs self-coupling $\lambda(E)$
     can be driven to a negative value and the Higgs potential
     becomes unbounded from below.
     A typical remedy for the situation is new physics
     which should appear before the crucial energy $\Lambda$ 
     where $\lambda$ reaches a zero value:
\begin{equation}\label{ST-limit}
\Lambda \le v\exp\left( 4\pi^2 \frac{M^2_h}{3\lambda_t^4 \, v^2}\right) 
         =  v\exp\left( 4\pi^2 
          \frac{2\lambda}{3\lambda_t^4}\right)  .
\end{equation}
     Here $M_h$ and $\lambda_t$ 
     are the Higgs mass and top
     Yukawa coupling at the weak scale.
     For a fixed value of the SM cut-off $\Lambda$ this relation 
     gives a lower (stability) bound on the Higgs boson mass
     and the self-coupling $\lambda$.
     For the first time, such a lower bound for the Higgs boson mass
     was obtained to be 3.7~GeV 
\cite{Linde:1977mm,Weinberg:1976pe}.

\subsection{\boldmath Some words about Higgs effective potential}
     A convenient tool for studying electroweak symmetry breaking
     in the SM is the analysis of the effective Coleman-Weinberg potential
	\cite{Coleman:1973jx,Gunion:1989we}. 
	Roughly speaking, this effective potential of the Higgs field 
	$V_{\rm eff}(\phi)$ 
	contrary to the classical potential 
	$V(\phi)$, given by (\ref{Lagr-scalar}), 
	takes into account the quantum corrections 
	to the energy density 
	of the field $\phi$. 
	The absolute minimum of the potential $V_{\rm eff}(\phi)$ 
	corresponds to the true vacuum state of the theory.

	In general, calculation of $V_{\rm eff}(\phi)$ is not 
	an easy task. 
	One usually turns to the loop expansion in order to obtain some 
	useful approximation for $V_{\rm eff}(\phi)$. 
	In the leading order approximation 
	$V_{\rm eff}(\phi)$ coincides with the classical 
	(so-called tree-level) potential $V(\phi)$. 
	The one-loop contributions arise due to 
	interactions of the Higgs field $\phi$ with the other
	fields of the theory. 
	With every bosonic (fermionic) field 
	which couples to the Higgs boson, the loop contribution 
	of the form 
\begin{equation}\label{CW-oneloop}
\Delta V (\phi) = 
\int \frac{d^4\,k}{2\, (2\pi)^4} \mathrm{STr}\ln \left(k^2 + M^2(\phi)\right)
\end{equation}
	is associated. 
	Here the supertrace counts positively (negatively)
	the number of degrees of freedom of  
	the corresponding particle and 
	$M^2(\phi)$ denotes 
	the field-depended mass that usually has
	the form 
\begin{equation} \label{FD-Mass}
	M^2(\phi) = \kappa \, \phi^2 + \kappa'.
\end{equation}
	Momentum integral (\ref{CW-oneloop})
	can be evaluated in the theory 
	defined with a momentum cut-off $\Lambda$
\begin{equation}\label{CW-oneloop-res}
\Delta V(\phi) = 
	- \frac{\Lambda^4}{128 \pi^2} \mathrm{STr} 1	
	+ \frac{\Lambda^2}{32 \pi^2} \mathrm{STr} M^2(\phi)
	+ \frac{1}{64 \pi^2} \mathrm{STr} M^4(\phi) 
	\left( \ln \frac{M^2(\phi)}{\Lambda^2} - \frac{1}{2} \right),
\end{equation}
	where all the terms that vanish in the limit 
	$\Lambda \to \infty$ are neglected.
	The first term in (\ref{CW-oneloop-res})
	contributes to the vacuum energy (cosmological constant).
	From 
(\ref{FD-Mass}) and the second term of  
(\ref{CW-oneloop-res})
	one can deduce the quadratic dependence of the Higgs mass 
	on the cut-off momentum (see the next section). 
	The last term in (\ref{CW-oneloop-res}) gives rise 
	to the effective Higgs boson self-couplings. 
	Clearly, the nonzero effective Higgs self-interactions 
	will be generated even 
	if one sets the initial self-coupling constant $\lambda$ to zero. 
	However, as was pointed out in the previous section, 
	the negative top-quark contribution in this case 
	($\lambda \approx 0$)
	will dominate 
	and will make the potential unbounded from below.

	It should be noted that one usually uses the renormalized form
	of potential 
(\ref{CW-oneloop}). In this form
	there is no (nonanalytical) dependence of the result 
	on the regularization parameter,
	e.g., on the cut-off $\Lambda$.
	However, one needs to introduce an auxiliary normalization scale $M$
	in order to define renormalized parameters and fields. 
	Independence of the physics on the mass scale $M$ can be used
	to extend the domain of the validity of the one-loop approximation
	by means of renormalization group method 
	(see, e.g. \cite{Quiros:1997vk}).

	Detailed study of the effective potential can be found, e.g., in 
\cite{Sher:1988mj,Casas:1996aq}. 
        For completeness in 
Table~\ref{tab:SM-FD-Mass} 
	we present the values of $\kappa$ and $\kappa'$ from 
(\ref{FD-Mass}) together with the number 
	of degrees of freedom $n$ for particles of the SM that give
	a dominant contribution to the effective potential.
\begin{table}[ht]
\centering
\begin{tabular}{|c|c|c|c|}
\hline
~~Particle~~& $\kappa$          &~~$\kappa'$~~&~~~n~~~~\\ \hline
$W^\pm$ & $g_2^2/4$         & 0 & $2 \times 3$  \\
$Z$     &~~~$(g_2^2+g_1^2)/4$~~~& 0 & 3 \\
$t$     & $\lambda_t^2/2$   & 0 & $4 \times 3$ \\
$h$     & $3 \lambda$       & $m^2$ & 1 \\
$\zeta$     & $ \lambda $       & $m^2$ & $3 \times 1$ \\	
\hline
\end{tabular}
\caption{Field-dependent masses of the SM particles 
$M^2(\phi) = \kappa \phi^2 + \kappa'$  together with the corresponding
numbers of degrees of freedom $n$. 
Massive vector bosons $W^+$, $W^-$, and $Z$ have 3 polarizations. 
The top quark $t$ besides usual 4 fermionic degrees of freedom has 
extra 3 color degrees of freedom. 
The Higgs field $h$ and 3 Nambu-Goldstone bosons $\zeta^{\pm,0}$ 
are scalars and have only one degree of freedom. 
When $\phi = v$, one obtains usual expressions 
for the tree-level masses.}
\label{tab:SM-FD-Mass}
\end{table}

\subsection{Quantum instability of the Higgs mass in the SM}
      There is also {\em quantum level}\/ 
      instability of the Higgs physics in the SM.
      The above-mentioned radiative corrections 
      are actually very severe for the
      (tachionic) mass term of the Higgs potential, since it 
      reveals itself to be highly dependent on the ultra-violet (UV) physics
      cut-off $\Lambda$
      (which leads to the so-called hierarchy problem)
\cite{Grojean:2005ud}.
      The one-loop (quantum) contributions to the 
      calculated SM Higgs boson mass $M_h$ can be 
      presented as  
\cite{Grojean:2005ud,Martin:1997ns}
\begin{equation}\label{Look-D-M_h}
\delta M^2_h = 
\left(
     \frac{9}{4}g^2_2 +\frac{3}{4}g^2_1 - 6\lambda^2_t +6 \lambda
\right) \frac{\Lambda^2}{32\pi^2}.
\end{equation}
       The SM (only) particles give unnaturally large corrections
       to the Higgs mass, they destabilize the Higgs vacuum
       expectation value $v$ and
       tend to push it towards the UV cut-off $\Lambda$ of the SM.

      The triviality and instability problems of the
      Higgs quartic self-coupling $\lambda$ can be avoided if one can find
      symmetry which can relate $\lambda$ with gauge
      coupling(s), for instance, in the form $\lambda = g^2$. 
      In this case, $\lambda$ would automatically possess
      the good UV asymptotically free behavior of
      the gauge coupling.
      Such a situation is realized 
      in the supersymmetric (SUSY) theories.
      Just for illustration, one can have a look at 
      the SUSY neutral scalar Higgs 
      potential from 
\cite{Martin:1997ns}
\begin{eqnarray}\label{ST-SUSY-H}
V (H_u^0, H_d^0)
&=&
(|\mu|^2 + m^2_{H_u}) |H_u^0|^2 + (|\mu|^2 + m^2_{H_d}) |H_d^0|^2
- (b\, H_u^0 H_d^0 + {\rm h.c.})
\nonumber \\ && 
+ {1\over 8} (g_2^2 + g_1^{2})(|H_u^0|^2 - |H_d^0|^2 )^2 .
\end{eqnarray}
     Here $H_u^0$ and $H_d^0$ are neutral components of the
     relevant SUSY Higgs fields,  
     $\mu$, $m^{}_{H_{d,u}}$, and $b$ are some SUSY parameters, 
     and $g_1$, $g_2$ weak gauge couplings.

    Some other 
    reviews of the SM Higgs constraints  can be found, for example, in 
\cite{Hambye:1997ax,Kolda:2000wi}. 
    In 
\cite{Hambye:1997ax}, 
    the two-loop Higgs mass upper bounds were reanalyzed.
    It was shown that the 
    previous results for a cut-off scale $\Lambda\approx$ few TeV 
    are too stringent.
    For $\Lambda=10^{19}$ GeV it was found that $M_h < 180 \pm 4\pm 5$ GeV,
    where the first error gives  theoretical uncertainty
    and the second error reflects 
    the experimental uncertainty in the top quark mass.
    A SM Higgs mass in the range of 160 to 170 GeV will
    certainly allow for a perturbative and well-behaved SM up to
    the Planck-mass scale $\Lambda_{\rm Pl}\simeq 10^{19}$ GeV,
    with no need for new physics to be set in below this scale
\cite{Hambye:1997ax}.

      The correlation between the Higgs mass of the
      SM and the scale at which the new physics is
      expected to occur is studied in 
\cite{Kolda:2000wi}. 
      Particular attention was paid 
      to the constraint imposed by the
      absence of the fine-tuning in the Higgs mass parameter
      (the Veltman condition).
      The Veltman condition (compare with the second term in  
      (\ref{CW-oneloop-res}) and eq.~(\ref{Look-D-M_h}))
$$
\frac{\Lambda^2}{32\pi^2}{\rm STr} M^2(\phi) =0 
$$    
      cancels the 
      1-loop quadratically divergent contributions to 
      the effective potential.
	Considering the coefficient in front of the $\phi^2$ term 
      in the above equation one can deduce that 
$3(2M^2_W+M^2_Z+M^2_h-4M^2_t)=0$ which results in the relation
$M_h=(317\pm 11)$~GeV for the Higgs mass
\cite{Kolda:2000wi}.
      It was found that the fine-tuning condition
      places a significant constraint also 
      on the new physics scale for
      the Higgs mass range 100~GeV$ < M_h <$~200~GeV mostly
      unconstrained by the classic constraints
      of unitarity, triviality, and vacuum stability
\cite{Kolda:2000wi}. 
 
       In fact, all above-mentioned 
       constraints (triviality, unitarity, stability, etc) 
       on the Higgs mass are tightly connected
       with the scale $\Lambda$, where one can, or should
       expect the new physics phenomena to occur
(see, for example \cite{Djouadi:2005gi}).

\subsection{Higgs vacuum and Cosmology}
       Closing this section we touch a less important question arising 
       in the cosmology due to 
       the Higgs mechanism and the nonzero vacuum expectation value
       of the scalar field. 
       From sections \ref{simplest} and \ref{HM-CSF}
       one concludes that 
       the vacuum state corresponds to  
       the negative value of the scalar potential in the minimum  
(\ref{Lagr-scalar-sigma}): 
$\displaystyle V(v)_{\min} = -\frac{\lambda v^4}{4}$. 
        With $v\simeq 246~$GeV, one has 
$\displaystyle V(v)_{\min} \simeq  - 10^9\lambda$~GeV$^4$ and this 
        is the contribution to vacuum energy of the Universe 
	due to spontaneous symmetry breaking.
	It is known from the cosmological observations that 
	the total energy density 
	of the Universe is rather small.
	It is at a level of $10^{-4}$~GeV/cm$^3$.
	Using the relation 
	1~GeV$^3=1.3 \times 10^{41}~$cm$^{-3}$
	(when $c=\hbar=1$), one has 
	a huge value	 
$\displaystyle V(v)_{\min} \simeq  - 10^{50}\lambda$~GeV$/$cm$^3$	
	for the Higgs contribution to the vacuum. 
        For reasonable $\lambda \simeq 0.1~(0.001)$ one obtains
	the contribution which $10^{54}~(10^{52})$ times larger than 
	the total energy of the Universe. 
	One solution to avoid this horrible situation is very simple. 
	It is sufficient to add a constant term 
	(bare cosmological constant) to the 
	potential and forget about the discrepancy.
	For example, the scalar potential can be taken in the
	form 
\cite{Okun:1990aa,Okun:1982ap,Grojean:2005ud}
$$V(\phi)=\frac{\lambda}{4}\left(\phi^2-v^2\right)^2
$$ 
        which has in its minimum $V(\phi=v)_{\min}=0$, by construction.
	Nevertheless, if one takes the problem more seriously,
	then to reach agreement of the 
	Higgs vacuum energy with its Universe value,	 
	one should adjust the constant 
	with accuracy $10^{-54}$ or so.
	The task looks completely meaningless, and reflects 
	a famous problem of the Einstein's cosmological constant.
	Furthermore, including gravity into consideration 
	one should take into account this above-mentioned Higgs vacuum term
	which strongly changes the space-time geometry
\cite{Kane:1987gb}.	 
        This observation gives one an almost obvious hint that the 
	spontaneous symmetry breaking Higgs mechanism
	has to be tightly connected with gravity. 

\section{Other ways to electro-weak symmetry breaking}
     Following Haber
\cite{Haber:2004tm},
    a very short list of other possible ways 
    for electro-weak symmetry breaking (EWSB) and 
    particle mass generation 
    is given in this section.

    In addition to the scalar dynamics of the SM, there have been many
    theories to explain the mechanism of EWSB.
    Some theories employ weakly-coupled scalar dynamics, 
    while others employ strongly-coupled dynamics 
    of a new sector of particles.  
    The motivation of nearly all proposed theories of
    EWSB beyond the SM is to address theoretical problems of
    naturalness and hierarchy.  
     The light Higgs bosons of 
{\em Little Higgs models}
\cite{Schmaltz:2005ky} 
     are nearly indistinguishable
     from the elementary Higgs scalars of the 
     weakly-coupled EWSB theories.  
     However, the new physics phenomena must enter here 
     near the TeV scale to cancel out the one-loop quadratic
     sensitivity of the theory to the ultraviolet scale.
     These theories have an implicit cut off of about 10~TeV,
     above which one would need to find their ultraviolet completions.  
     The {\em extra-dimensional theories of EWSB} 
\cite{Quiros:2003gg}
      lead to new models of the EWSB dynamics, including the
      so-called ``Higgsless'' models 
\cite{Nomura:2003du,Simmons:2006iw}
      in which there is no light Higgs scalar in the spectrum.  
      Such models also require an ultraviolet completion
      at a scale characterized by the inverse radius of extra dimension.
      Models of {\em strongly-coupled EWSB sectors}  
\cite{Hill:2002ap}
      include technicolor models, composite Higgs models
      of various kinds, top-quark condensate models, etc.

     The new physics beyond the SM can be of two types --- decoupling 
\cite{Gunion:2002zf} or non-decoupling.  
    The virtual effects of ``decoupling'' physics beyond the SM
    typically scale as
    $m_Z^2/M^2$, where $M$ is a scale characteristic of the new physics.
    Examples of this type include ``low-energy'' supersymmetric theories
    with soft-supersymmetry-breaking masses of $\mathcal{O}(M)$.  
In
contrast, some of the virtual effects of ``non-decoupling'' physics 
do not vanish as the characteristic scale $M\to\infty$.  A theory
with a fourth generation fermion and technicolor models are examples
of this type.  Clearly, the success of the SM electroweak fit places
stronger restrictions on non-decoupling new physics.  Nevertheless,
some interesting constraints on decoupling physics can also be
obtained.  For example, even in theories of the new physics that exhibit
decoupling, the scale $M$ must be somewhat separated from the scale $m_Z$ 
(to avoid a conflict with the SM electroweak fit).  This leads to a
tension with the requirements of naturalness which has been called
the ``little hierarchy problem'' 
\cite{Cheng:2003ju} in the literature.

\section{Conclusion}
      The Higgs mechanism in the framework of the Standard Model is reviewed. 
      The discussions of the Higgs self-coupling  $\lambda$ parameter and the 
      bounds for the Higgs boson mass are presented in detail.
      In particular, the unitarity, triviality and stability 
      constraints on $\lambda$ are discussed.
      The generation of the finite value for the $\lambda$ parameter 
      due to quantum 
      corrections via the effective potential is illustrated.
      A simple case with both the top-Higgs Yukawa coupling and 
      the Higgs self-coupling $\lambda$ equal to 1 is considered
      and the top quark mass to be 174.1 GeV and
      Higgs boson mass to be 348.2 GeV are predicted.
      A short list of other ways 
      for the electro-weak symmetry breaking and 
      the particle mass generation beyond the Standard Model is given.

      Finally, following L.B.~Okun 
\cite{Okun:1990aa,Okun:1982ap,Okun:1984aa}, we would like to stress
      that it looks like that there is no way out of scalar particles.
      They are inevitable.
      With these scalars the most fundamental problems of  
      modern particle physics are connected,  
      in particular, they are the problem of particle mass generation, 
      the cosmological inflation, and the dark energy.
      While vector fields describe the dynamics of
      interactions, the scalar 
      fields are responsible for inertia.
      While vector fields are results of local symmetry, 
      the scalar fields carry 
      the symmetry breaking function, 
      the function of the 
      same level of importance.
      Therefore, the most important task of current physics research
      is to discover scalar particles and study their
      properties
\cite{Okun:1984aa}.

\smallskip
       This work was supported 
       by the Russian Foundation for Basic Research (grants 06--02--04003
       and 05--02--17603).
       The authors thank 
       Prof. J.A.~Budagov and D.I.~Kazakov 
       for fruitful collaboration and 
       for useful discussions.

\section{Appendices}
\subsection{The Standard Model before electroweak symmetry breaking} 
\label{App-SM}
    The electroweak theory of Glashow, Weinberg, and Salam 
\cite{Glashow:1961tr,Weinberg:1967tq,Salam:1969aa} 
    describes the electromagnetic and weak interactions between
    quarks and leptons. 
    It is the Yang--Mills theory 
\cite{Yang:1954ek} constructed  
    on the symmetry group SU(2)$_{\rm L}\times $U(1)$_{\rm Y}$. 
    Combined with quantum chromodynamical (QCD) 
    SU(3)$_{\rm C}$ gauge theory 
    of strong interactions 
\cite{Gell-Mann:1964nj,Fritzsch:1973pi,Gross:1973id,Politzer:1973fx}, 
    it has the name of the Standard Model (SM). 
    Pattern of interactions (governed by underlying symmetries and
    given in the form of Lagrangians)
    and the field content are both two main ingredients of the SM.
    The model (before the electroweak
    symmetry breaking) has two kinds of fields. 
    First, there are three generations of
    left-handed and right-handed 
    chiral (matter fields) quarks and leptons, 
    $f_{\rm L,R} =\frac{1}{2} (1 \mp \gamma_5)f$. 
    The left-handed fermions are in weak isodoublets
    (with the third component of the weak isospin 
    $T^3_{f_{\rm L}}= \pm \frac{1}{2}$), 
    while the right-handed fermions are 
    weak isosinglets
    (with  $T^3_{f_{\rm R}}= 0$) 
\begin{eqnarray} 
\begin{array}{l} 
L_1=\left(\!\begin{array}{c}\nu_{e\rm L}\\e^-_{\rm L}\end{array}\!\right),\  
e_{\rm R_1}=e^-_{\rm R}, \qquad 
Q_1=\left(\!\begin{array}{c}u_{\rm L}   \\  d_{\rm L}\end{array}\!\right),\ 
u_{\rm R_1}=u_{\rm R}, \quad d_{\rm R_1}=d_{\rm R};  
\\[6pt] 
L_2=\left(\!\begin{array}{c}\nu_{\mu\rm L}\\\mu^-_{\rm L}\end{array}\!\right),\
e_{\rm R_2} =\mu^-_{\rm R},  \qquad
Q_2= \left(\!\begin{array}{c}c_{\rm L}   \\  s_{\rm L}\end{array}\!\right),\ 
u_{\rm R_2}=c_{\rm R},\quad  d_{\rm R_2} = s_{\rm R}; 
\\[6pt] 
L_3=\left(\!\begin{array}{c}\nu_{\tau\rm L}
     \\\tau^-_{\rm L}\end{array}\!\right),\    
e_{\rm R_3}=\tau^-_{\rm R}, \qquad 
Q_3=\left(\!\begin{array}{c}t_{\rm L}\\b_{\rm L}\end{array}\right),\ 
u_{\rm R_3}=t_{\rm R},\quad  d_{\rm R_3}= b_{\rm R}. 
\end{array}
\end{eqnarray}
     The fermion hypercharge
\begin{equation}\label{Hypercharge}
Y_f=2Q_f-2T_f^3,
\end{equation} 
     defined in terms of the third component of the weak isospin $T_f^3$ 
     and the electric charge $Q_f$ in units of the proton charge 
     $+e$ is given by ({\small $i$=1,2,3})
\begin{eqnarray}\label{Hypercharges}
Y_{L_i}=-1, \quad Y_{e_{\rm R_i}}=-2, \quad  
Y_{Q_i}=\frac{1}{3}, \quad 
Y_{u_{\rm R_i}}= \frac{4}{3}, \quad  Y_{d_{\rm R_i}}= -\frac{2}{3}. 
\end{eqnarray}
          Moreover, the quarks are triplets under the 
	  ${\rm SU(3)_C}$ group, while leptons
	  are color singlets. This leads to the relation 
$  \sum_f Y_f =\sum_f Q_f=0$ 
       which ensures the cancellation of chiral anomalies 
       within each generation, thus preserving 
       the renormalizability of the electroweak theory
       (see, for example \cite{Djouadi:2005gi}). 

    Second, there are gauge fields corresponding to spin-one bosons
    that mediate interactions. 
    In the electroweak sector, one has the field 
    $B_\mu$ which corresponds to the generator $Y$ 
    of the U(1)$_{\rm Y}$ group and the three fields 
    $W^{1,2,3}_\mu$ which correspond to the generators 
    $T_{i} = \frac{1}{2} \tau_i$ 
    of the SU(2)$_{\rm L}$ group
    with the commutation relations between these generators 
\begin{eqnarray}
[T^i,T^j]=i\epsilon^{ijk} T_k \quad {\rm and} \quad [Y, Y]=0. 
\end{eqnarray}
    Here $\epsilon^{ijk}$ is the antisymmetric tensor and 
    non-commuting $2 \times 2$ 
    Pauli matrices have their standard form 
\begin{eqnarray}\label{Pauli-matrices}
\tau_1= \left( \begin{array}{cc} 0 & 1 \\ 1 & 0 \end{array} \right) \, , \ 
\tau_2= \left( \begin{array}{cc} 0 & -i \\ i & 0 \end{array} \right) \, , \ 
\tau_3= \left( \begin{array}{cc} 1 & 0 \\ 0 & -1 \end{array} \right).
\end{eqnarray}
     There is an octet of gluon fields $G_\mu^{a}$ 
     in the strong interaction sector. 
     The gluon octet correspond
     to 8 generators of the ${\rm SU(3)_C}$ group 
     which obey the relations
$$ 
[T^a,T^b]=if^{abc} T_c,  \quad {\rm with} \quad 
{\rm Tr}[T^a T^b]= \frac12 \delta_{ab} 
$$ 
     where the tensor $f^{abc}$ is for the structure constants of the 
     ${\rm SU(3)_C}$ group. 
    The field strengths are given by
\begin{eqnarray*}
G_{\mu \nu}^a &=& \partial_\mu G_\nu^a -\partial_\nu G_\mu^a +g_s \, 
f^{abc} G^b_\mu G^c_\nu, \non \\
W_{\mu \nu}^a &=& \partial_\mu W_\nu^a -\partial_\nu W_\mu^a +g_2 \, 
\epsilon^{abc} W^b_\mu W^c_\nu, \non \\ 
B_{\mu \nu} &=& \partial_\mu B_\nu -\partial_\nu B_\mu, 
\end{eqnarray*}
    where $g_s$, $g_2$ and $g_1$ are, respectively, the coupling constants of 
    ${\rm SU(3)_C}$,  ${\rm SU(2)_L}$ and  ${\rm U(1)_Y}$.  
	There are triple 
$~ig_i\,{\rm Tr} (\partial_\nu V_\mu-\partial_\mu V_\nu)[V_\mu,V_\nu]$~
        and quartic $~\frac12 g_i^2 \, {\rm Tr} [V_\mu,V_\nu]^2$~
	self-interactions between 
	non-Abelian gauge fields 
	$V_\mu \equiv W_\mu$ (SU(2) group) or $G_\mu$ (SU(3) group). 
      The matter fields $\psi$ are       coupled to the gauge 
      fields through the covariant derivative
\begin{eqnarray}
D_{\mu} \psi = \left( \partial_\mu -ig_s T_a G^a_\mu - ig_2 T_i W^i_\mu 
-i g_1 \frac{Y_q}{2} B_\mu \right) \psi  
\label{CovariantDerivative}
\end{eqnarray}
      which leads to unique couplings between the fermion and 
      gauge fields 
$-g_i \overline \psi V_\mu \gamma^\mu \psi$. 

       The SM Lagrangian {\em before electroweak symmetry breaking}\/
       (without mass terms for fermions and gauge bosons) 
       is given by 
\begin{eqnarray}
\label{smlagrangian}
{\cal L}_{\rm SM}
&=& -\frac{1}{4} G_{\mu \nu}^a G^{\mu \nu}_a 
-\frac{1}{4} W_{\mu \nu}^a W^{\mu \nu}_a -\frac{1}{4} 
B_{\mu \nu}B^{\mu \nu} \\ 
&& + \bar{L_i}\, i D_\mu \gamma^\mu \, L_i + \bar{e}_{Ri} \, i D_\mu 
\gamma^\mu \, e_{R_i} \ 
+ \bar{Q_i}\, i D_\mu \gamma^\mu \, Q_i + \bar{u}_{Ri} \, i D_\mu 
\gamma^\mu \, u_{R_i} \ + \bar{d}_{Ri} \, i D_\mu \gamma^\mu \, d_{R_i}. \non 
\end{eqnarray}
     This Lagrangian is invariant under local 
     ${\rm SU(3)_C \times SU(2)_L \times U(1)_Y}$ gauge 
     transformations for fermion and gauge fields. 
     For instance, in the electroweak sector one has  
\begin{eqnarray}
L(x) \to L'(x)=e^{i\alpha_i(x) T^i + i \beta(x)Y } L(x), \quad
R(x) \to R'(x)=e^{i \beta (x) Y} R(x), \non \\
\vec{W}_\mu (x) \to \vec{W_\mu}(x) -\frac{1}{g_2} 
\partial_\mu \vec{\alpha}(x)- 
\vec{\alpha}(x) \times \vec{W}_\mu(x), \quad 
B_\mu(x) \to B_\mu(x) -  \frac{1}
{g_1} \partial_\mu \beta (x).
\end{eqnarray}
       Up to now, the gauge  and  fermion fields have been kept 
       massless.
       In the case of strong interactions, the gluons are indeed 
       massless particles while mass terms of the form  
       $-m_q\overline{\psi}\psi$ can be generated for
       the colored quarks 
       in an SU(3) gauge invariant way. 
       In the electroweak sector is it impossible to do so.
       `By-hand'' 
       incorporation of mass terms for gauge bosons and  
       fermions leads to a 
       breakdown of the local ${\rm SU(2)_L\times U(1)_Y}$ gauge invariance. 
       Only due to 
       spontaneous symmetry breaking 
       one can generate the gauge boson and
       the fermion masses without violating  
       SU(2)$\times$U(1) gauge invariance. 

\subsection*{The Standard Model after the electroweak symmetry breaking} 
      The basis of the Standard Model is 
      the SU(3)$\times$SU(2)$\times$U(1) gauge invariance 
      together with the 
      electroweak symmetry breaking (Higgs) mechanism
(see, for example \cite{Djouadi:2005gi}). 
       The Higgs mechanism of 
       spontaneous symmetry breaking and mass generation
       in the SM is given in detail in section
\ref{HM-SM}.
       Below only some most important relations
       following from the Higgs mechanism are collected.

      The scalar field 
      vacuum expectation value $v$ is fixed in terms of the $W$ 
      boson mass $M_W$ and the Fermi constant $G_{\rm F}$ 
\begin{eqnarray}
M_W=\frac{g_2v}{2}=
\left(\frac{\sqrt{2}g_2^2}{8 G_{\rm F}} \right)^{1/2} \qquad 
{\rm and }\qquad
v= \frac{1}{(\sqrt{2} G_{\rm F}
)^{1/2} } \simeq 246~{\rm GeV}.
\label{MW-vs-v} 
\end{eqnarray}
     The muon decay lifetime is very precisely  measured experimentally. 
     It is directly related 
     to the Fermi coupling constant by means of the following 
     relation which includes QED corrections 
\cite{Behrends:1955mb,Kinoshita:1958ru,vanRitbergen:1998yd}
\begin{eqnarray}\label{Mu-width}
\frac{1}{\tau_\mu} &=& \frac{G_{\rm F}^2 m_\mu^5}{192 \pi^3} 
\left( 1- \frac{8m_e^2}{m_\mu^2} 
\right) \left[1+ 1.810 \frac{\alpha}{\pi} 
+  (6.701 \pm 0.002) \left( \frac{\alpha} {\pi} \right)^2 \right]
\end{eqnarray}
       where $m_e$ and $m_\mu$ are the electron and muon masses 
       and $\alpha$ is the fine-structure constant.
       From 
(\ref{Mu-width}) one has the precise value of the Fermi constant
\cite{Yao:2006px} 
\begin{eqnarray}\label{GFermi}
G_{\rm F} = (1.16637 \pm 0.00001) \cdot 10^{-5}~{\rm GeV}^{-2} .
 \end{eqnarray}    
      In the SM, the muon decay occurs through gauge interactions 
      mediated by $W$ boson exchange and, therefore, 
      one obtains a relation between the $W$, $Z$  masses,
      $\alpha$ and $G_{\rm F}$ 
\begin{eqnarray}
\frac{G_{\rm F}}{\sqrt{2}} = \frac{g_2}{2\sqrt{2}} \cdot \frac{1}{M_W^2}
\cdot \frac{g_2}{2\sqrt{2}} = \frac{ \pi \alpha}{2 M_W^2 s_W^2} 
= \frac{ \pi \alpha}{2 M_W^2 (1- M_W^2/M_Z^2)},
\qquad 
\frac{g^2_2}{4}=\frac{\pi\alpha}{\sin^2\theta}. 
\label{App-SM-alpha}
\end{eqnarray}
       From these relations one can derive formula
(\ref{MW-vs-v}). 
       The gauge field rotation to the physical gauge bosons 
       (mass eigenstates), given by relation 
(\ref{ZA-fields}), defines the electroweak mixing angle
       $\sin \theta_W$
      which can also 
      be written in terms of the $W$ and $Z$ boson masses
\begin{eqnarray}\label{sinW}
\sin \theta_W = \frac{g_1}{\sqrt{g_1^2 + g_2^2}} = \frac{e}{g_2},
\qquad
\sin^2 \theta_W 
= 1- \frac{M_W^2}{M_Z^2}.
\label{sw-definition}
\end{eqnarray} 
      Using the fermionic part of the SM Lagrangian
(\ref{smlagrangian}), 
      written in terms of the new fields and writing explicitly 
      the covariant derivative one obtains 
\begin{eqnarray}
{\cal L}_{\rm NC} &=& e J_\mu^{A} A^\mu + \frac{g_2}{\cos\theta_W} 
J_\mu^Z Z^\mu, \non \\
{\cal L}_{\rm CC} &=& \frac{g_2} {\sqrt{2}} (J_\mu^+ W^{+\mu}
+ J_\mu^- W^{-\mu}) 
\end{eqnarray}
      for the neutral and charged current parts, respectively. 
      The currents $J_\mu$ are then given by 
\begin{eqnarray}
J_\mu^A &=& Q_f \bar{f}\gamma_\mu f, \non \\  
J_\mu^Z &=& \frac{1}{4}  \bar{f}
\gamma_\mu [ (2T_f^3 - 4 Q_f \sin^2 \theta_W) - \gamma_5 (2T^3_f) ]f, \non \\ 
J_\mu^+ &=& \frac{1}{2} \bar{f}_u \gamma_\mu (1- \gamma_5) f_d
\end{eqnarray}
      where $f_u (f_d)$ is the up-type (down-type) 
      fermion of isospin $+(-)\frac{1}{2}$
\cite{Djouadi:2005gi}. 

      In terms of the electric charge $Q_f$ of the fermion $f$ 
      and with the left-handed weak 
      isospin of the fermion $T_f^3 = \pm \frac{1}{2}$ 
      and the weak mixing angle 
      $s_W^2=1-c_W^2 \equiv \sin^2 \theta_W$, one can write the vector 
      and axial vector couplings of the fermion $f$ to the $Z$ boson 
\begin{eqnarray}
v_f = \frac{ \hat{v}_f} { 4 s_W c_W} = 
\frac{ 2T_f^3 -4 Q_f s_W^2}{ 4 s_W c_W}, \ \ 
a_f = \frac{ \hat{a}_f} { 4 s_W c_W} = 
\frac{ 2T_f^3}{ 4 s_W c_W}
\label{Zffcouplings}
\end{eqnarray}
        where we also defined the reduced $Z f\bar f$ couplings 
	$\hat{v}_f, \hat{a}_f$.
	In the case of the $W$ boson, its vector and axial-vector 
	couplings to fermions are simply 
\begin{eqnarray}
v_f = a_f = \frac{1}{2 \sqrt{2}s_W} = \frac{\hat a_f}{4 s_W} = 
\frac{\hat{v}_f}{4 s_W} .
\label{Wffcouplings}
\end{eqnarray}
       These results are only valid in the one-family approximation. 
       While the extension to three families is straightforward 
       for neutral currents, there is a complication in the 
       case of the charged currents
       due to the fact that the current eigenstates 
       for quarks $q'$ are not identical to the mass eigenstates $q$. 
       If we start by $u$-type quarks being mass eigenstates, 
       in the down-type quark sector, the two sets are 
       connected by a unitary transformation 
\begin{eqnarray}
(d',s',b') = V (d,s,b)
\end{eqnarray}
        where $V$ is the $3\!\times\!3$ 
	Cabibbo--Kobayashi--Maskawa (CKM) matrix. 
	The unitarity of $V$ insures that the neutral currents 
	are diagonal in both the bases. 
	This is the GIM mechanism 
	which ensures a natural absence of flavor changing
	neutral currents (FCNC) at the tree level in the SM. 
	For leptons, the mass and current  eigenstates coincide, 
	since in the SM the neutrinos are assumed to
	be  massless, which is an excellent  approximation in most purposes.

	Note that the relative strength of the charged and 
	neutral currents, $J^\mu_Z J_{\mu Z}/J^{ \mu +} J_\mu^-$ 
	can be measured by the parameter $\rho$ 
	which, using the previous formulas, is given by
\begin{eqnarray}
\rho=  \frac{M_W^2}{c_W^2 M_Z^2}
\label{rho-MWMZ}
\end{eqnarray}
       and is equal to unity in the SM due to 
(\ref{sw-definition}). 
       This is a direct consequence of the choice of the 
       representation of the Higgs field responsible for 
       breaking of the electroweak symmetry. 
       In a model which makes use of an arbitrary number of 
       Higgs multiplets $\Phi_i$ with isospin $I_i$,  
       the third component $I_i^3$ and vacuum expectation values 
       $v_i$, one obtains for this parameter
\begin{eqnarray}
\rho= \frac{\sum_i \left[ I_i (I_i+1) -(I_i^3)^2\right] v_i^2}
{2 \sum_i (I_i^3)^2 v_i^2}
\end{eqnarray}
       which is also unity for an arbitrary number of doublet 
       as well as singlet fields. 
         This is due to  the fact that in this case, the model has 
         custodial SU(2) global symmetry ($V(\Phi)$ in 
	 (\ref{SU2-ScalarL}) 
	 is invariant under global O(4)). 
	 In the SM, this symmetry is broken at the loop level 
	 when fermions of the same doublets have different masses and by the
	 hypercharge group. 

	 Finally, self-couplings among the gauge bosons are 
	 present in the SM as a consequence of the non-Abelian 
	 nature of the ${\rm SU(2)_L\times U(1)_Y}$ symmetry. 
	 These couplings are dictated by the structure of 
	 the symmetry group and, for instance, 
	 triple self-couplings among the $W$ and the 
	 $V=\gamma,Z$ bosons are given by
\begin{eqnarray}
{\cal L}_{WWV} = ig_{WWV} \left[ W^\dagger_{\mu \nu} W^{\mu} V^\nu - 
W^\dagger_{\mu} V_{\nu} W^{\mu \nu}  + W^\dagger_{\mu} W_{\nu} V^{\mu \nu}  
\right] \label{WWVcoupling}
\end{eqnarray}
          with $g_{WW\gamma}=e$ and $g_{WWZ}=e c_W/s_W$.
(for more details see, for example,~
\cite{Djouadi:2005gi}).

\subsection*{The SM particle masses}
     The top quark possessing the heaviest mass
     of currently known elementary particle 
     plays a very important role not only
     in the Higgs boson physics.
     The top quark was produced, for the first time, 
     at the Tevatron in the reaction $p\bar{p}
     \to q \bar{q}/ gg \to t\bar{t}$, and now it is under  
     permanent investigation at FNAL by the CDF and D\O\
     collaborations.
     In the SM, the top quark almost always 
     decays into a $b$ quark and a $W$ boson. 
     The width $t \to b W^+$ is given by 
(see, for example, \cite{Djouadi:2005gi,Dawson:2003uc,Kuhn:1996ug} 
and references therein)   
\begin{eqnarray}
\Gamma_t \simeq \Gamma (t \to b W^+) 
= \frac{G_{\rm F} m_t^3} {8\sqrt{2} \pi} 
|V_{tb}|^2 \, \left( 1- \frac{M_W^2}{m_t^2} \right)^2 
\left(1+2\frac{M_W^2}{m_t^2}\right) 
\left(1-2.72\frac{\alpha_s}{\pi}\right)+{\cal O}(\alpha_s^2,\alpha)
\end{eqnarray}
      and is of the order of $\Gamma_t \simeq 1.8$~GeV
       for $m_t \simeq 180$ GeV.
      Here $|V_{tb}|$ 
      is the top-bottom CKM matrix element and 
      $\alpha_s$ is the strong coupling constant. 
     The modern, average (over CDF and D\O), 
     mass value for the top quark is given by the PDG-2006
\cite{Yao:2006px}
\begin{equation}\label{top-mass}
m_t= 174.2 \pm 3.3~{\rm GeV}.
\end{equation}
	Given the experimental technique used to extract the top mass, these
	mass values should be taken as representing the top pole mass
\cite{Yao:2006px}, which corresponds to the pole in the top-quark
	propagator.  
	For an observable particle such as the electron 
	the pole mass is equal to its physical mass. 
	It is well known that the pole mass for the quark cannot be used to 
	arbitrarily high accuracy because of
	the nonperturbative infrared effects in QCD 
	which are of an order of 
	${\mathcal O}(\Lambda_{\rm QCD})$ 
(see, e.g., 
\cite{Yao:2006px}). 
	For the top quark mass one can 
	neglect this intrinsic ambiguity, since 
	the experimental errors are much higher.  
	However, for the $b$- and $c$-quarks 
	the ambiguity is significant,
	e.g., it is about 10\% for the $b$-quark pole mass, 
	so one usually has to define a more
	appropriate quark mass parameter. 
For example, 
	at high energies the 
so called ``short-distance'' running mass 
${\overline{m}}_{Q}(\mu)$
        is used, 
        since it is insensitive to any ``physics'' 
	at the distances larger than the scale of $1/\mu$. 
	Usually, one uses
	a modified minimal subtraction scheme 
	$\overline{\mathrm{MS}}$ to define 
	this quantity.
	In particular, for the running bottom and charm
	masses the PDG-2006 
\cite{Yao:2006px} gives
$$\overline{m}_b(\overline{m_b})=4.20 \pm 0.07~{\rm GeV}, 
\qquad
\overline{m}_c(\overline{m_c})=1.25 \pm 0.09~{\rm GeV}.  
$$ 
	For the strange quark one can typically use the value 
	$\overline{m}_s({\rm 1\, GeV})=0.2$ GeV.
	The masses of light $u,d$ quarks, being very small in comparison with
	the Higgs boson mass, are not given here.

   	In case one needs top quark running mass, one can use   
	the relation between the pole masses and the running 
   	masses 
\cite{Gray:1990yh,Melnikov:2000qh,Chetyrkin:1999ys,Chetyrkin:1999qi}
\begin{eqnarray} \label{run-pole}
{\overline{m}}_{Q}(m_{Q})= m_{Q} \, \bigg[ 1- \frac{4}{3}
\frac{\alpha_{s}(m_Q)}{\pi} + (1.0414 N_f - 14.3323) \frac{\alpha_s^2(m_Q)}
{\pi^2} \bigg] \non \\
 +(-0.65269 N_f^2 +26.9239 N_f -198.7068)\frac{\alpha_s^3(m_Q)}{\pi^2} \bigg]
\end{eqnarray}
   where $\alpha_s$ is the $\overline{\rm MS}$ 
   strong coupling constant evaluated 
   at the scale of the pole mass $\mu=m_Q$, and
   $N_f$ is the number of (active) quark flavors.   
	
   The evolution of $\overline{m}_Q$ 
   from the scale $\overline{m}_{Q}$ upward to 
a renormalization scale $\mu$ is 
\begin{eqnarray}
&& \hspace*{1.5cm}{\overline{m}}_{Q}\,(\mu )
={\overline{m}}_{Q}\,(\overline{m}_{Q})
\,\frac{c\,[\alpha_{s}\,(\mu)/\pi ]}
{c\, [\alpha_{s}\,(\overline{m}_{Q})/\pi ]}
\label{eq:msbarev} 
\end{eqnarray}
    with the function $c$, up to three-loop order, given by 
\cite{Gorishnii:1991zr,Gorishnii:1990zu,Chetyrkin:1997dh,Vermaseren:1997fq}
\begin{eqnarray*}
c(x)&=&\left(25x/6 \right)^{12/25} \,
       [1+1.014x+1.389\,x^{2} + 1.091\, x^3]
\hspace{1.0cm} \mbox{for} \hspace{.2cm} m_{c}\,<\mu\,<m_{b}\\
c(x)&=&\left(23x/6 \right)^{12/23} \,
       [1+1.175x+1.501\,x^{2} + 0.1725\, x^3]
\hspace{.8cm} \mbox{for} \hspace{.2cm} m_{b}\,<\mu \,< m_t \\
c(x)&=&\left(7x/2 \right)^{4/7} \,
       [1+1.398x+1.793\,x^{2} - 0.6834\, x^3]
\hspace{1.35cm} \mbox{for} \hspace{.2cm} m_{t}\,<\mu .
\end{eqnarray*}

     The values of the running $t$-, $b$-, and $c$-quark 
     masses at the scale $\mu = M^{}_Z= 91$~GeV are 
\cite{Chetyrkin:2000yt}
$$\overline{m}_t(M_Z) = 172.6~{\rm GeV}, \qquad
  \overline{m}_b(M_Z) = 2.87~{\rm GeV}, \qquad
  \overline{m}_c(M_Z) = 0.60~{\rm GeV}. 
$$
   The PDG-2006
\cite{Yao:2006px} masses of the charged leptons are the following: 
\begin{eqnarray*}
m_\tau=1.777~{\rm GeV},\quad 
m_\mu=0.1057~{\rm GeV},\quad 
m_e=0.511~{\rm MeV}. 
\end{eqnarray*}
    Finally, the masses and total decay
    widths of the two main gauge bosons are
\cite{Yao:2006px} 
\begin{eqnarray*}
M_Z=91.1876 \pm 0.0021 ~{\rm GeV}, 
&\quad& \Gamma_Z = 2.4952 \ \pm 0.0023  ~{\rm GeV}; 
\\ 
M_W=80.403 \pm 0.029~{\rm GeV}, 
&\quad& \Gamma_W= 2.141 \ \pm 0.041~{\rm GeV}. 
\end{eqnarray*}  

\subsection{\boldmath The Higgs mechanism with extra $\phi^3$-term}
\label{App-Phi3}
\begin{figure}[!h]
\begin{center} 
\epsfig{file=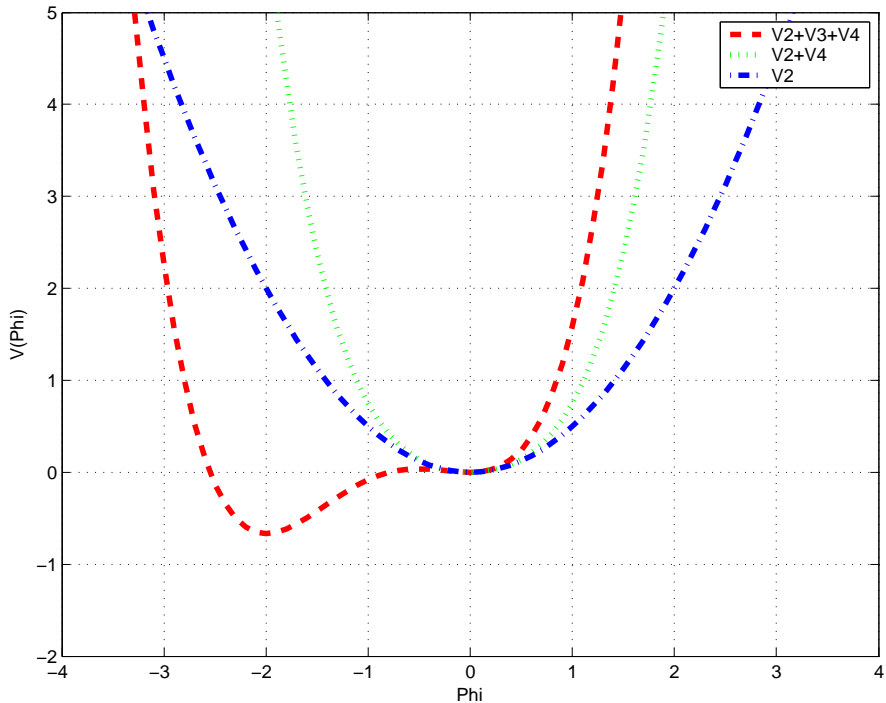,width=14.cm} 
\end{center}
\vspace*{-1cm}
\caption{
       The potential $V$ of the scalar field $\phi$ in 
       the case $\mu^2=1$, $\kappa=2.5$
       and $\lambda=1$.
       In this case there are 3 extrema at 
       $\phi^{(3)}=0$ with $V(\phi^{(3)}_0)=0$; 
       $\phi^{(2)}=-1/2$ with $V(\phi^{(3)}_0)\approx 0.0365$ 
       (local maximum) and 
       $\phi^{(1)}=-2$ with $V(\phi^{(3)}_0)=-0.6$ 
       (the only real minimum).
\label{V-Higgs-phi3}}
\end{figure}

      Consider a Lagrangian for scalar real field $\phi$ 
\begin{eqnarray}
\label{Lagr-scalar-phi3}
{\cal L}= \frac{1}{2} \partial_\mu \phi \, \partial^\mu \phi - V(\phi), 
\qquad {\rm where} \qquad 
V(\phi)= \frac{1}{2} \mu^2 \phi^2 + \frac{1}{3}\kappa \phi^3
+ \frac{1}{4}\lambda \phi^4.
\end{eqnarray}
       The Lagrangian 
(\ref{Lagr-scalar-phi3}) ``pretends to describe'' 
       a spin-zero particle of mass $\mu$ (with 
       cubic and quadric self-interactions).
      It is not invariant under the reflexion symmetry 
      $\phi\to -\phi$, since there is explicitly a cubic term.  
      Since the potential should be bounded from below
      the self-coupling $\lambda>0$. 
      However, contrary to the ordinary Higgs potential 
      (without the extra $\phi^3$-term),  
      also in the case when the mass term $\mu^2>0$, 
      the potential $V(\phi)$ can, in principle, be 
      negative for some $\phi$ due to the presence of 
      the cubic term. 
      This means that, for example, for 
      $\lambda>0$, $\mu^2>0$, $\kappa\ne 0$ 
      (with an arbitrary sign of $\kappa$)
      the potential can be negative for some $\phi$ and,
      therefore, can have a minimum
(see Fig.~\ref{V-Higgs-phi3}).
      In general, any 
      minimum of the potential can be obtained for 
      $\langle 0| \phi | 0 \rangle \equiv \phi_0$ 
     which solves a minimum (extremum) condition \
\begin{equation}\label{Phi3-minimum}
\partial V/\partial \phi =\phi(\mu^2 + \kappa \phi + \lambda \phi^2)=0.
\end{equation}
     There are 3 solutions. 
     One is obviously $\phi^{(3)}_0=0$, which gives $V(\phi^{(3)}_0)=0$.
     In principle, the two other solutions can be those 
     of the quadratic equation
$\lambda \phi^2 + \kappa \phi + \mu^2 =0$
(if $\kappa^2-4\lambda\mu^2>0$) 
$$
\phi^{(1,2)}_{0}
= \frac{-\kappa\pm \sqrt{\kappa^2-4\lambda\mu^2}}{2\lambda}.
$$
       It is obvious that (if $\kappa^2-4\lambda\mu^2>0$) 
       $|\phi^{(1)}_{0}| \ne |\phi^{(2)}_{0}|$ and in general
       $V(\phi^{(1)}_{0}) \ne V(\phi^{(2)}_{0})$.
       Therefore, only one real minimum for the potential $V$ exists
       (see 
Fig.~\ref{V-Higgs-phi3}).
       It is important to note that there are no any solutions 
   $\phi^{(1,2)}_{0}~$ if $~\kappa^2-4\lambda\mu^2<0~$
       and the true minimum stays at $\phi^{(3)}_0=0$.

       To simplify the problem, let us assume a ``massless'' scalar
       field $\phi$ with $\mu^2=0$.
       In this case, due to 
$\displaystyle
\phi^{(1,2)}_{0}= \frac{-\kappa\pm \sqrt{\kappa^2}}{2\lambda}
$
      one has only one nonzero $\phi_0$,
      say         
\begin{equation}\label{V-Higgs-v}
\phi^{(1)}_{0}= \frac{-\kappa}{\lambda} \equiv v, \quad
{\rm and} \quad
      \phi^{(3)}_{0}=\phi^{(2)}_{0}=0.
\end{equation}
     Here {\em the only}\/ 
     quantity $v \equiv \langle \, 0|\phi| 0 \,  \rangle$ 
     can be (as before) 
     called the vacuum expectation value (vev) of the scalar field
     $\phi$ and has a sign opposite to $\kappa$.
     Lagrangian 
(\ref{Lagr-scalar-phi3}) no longer describes a particle 
     with mass $\mu$ (or ever a massless particle when $\mu^2=0$). 
     To interpret correctly the theory, 
     one must expand around the real minimum $v$ 
     by defining the field 
     $\sigma$ as $\phi= v + \sigma$ and assuming that
     $\langle \, 0|\sigma| 0 \,  \rangle =0$. 
     In terms of the new field $\sigma$, the potential 
     $V(\phi)$ of 
(\ref{Lagr-scalar-phi3}) becomes ($\mu^2=0$ is assumed and 
$\kappa = - v\lambda$ is used)
\begin{eqnarray*} 
V(\phi) &=& V(\sigma) = 
            \frac{\kappa}{3}\phi^3 + \frac{\lambda}{4}\phi^4
         =  \left(\frac{\kappa}{3}+ \frac{\lambda}{4}\phi\right)
            \phi^3
         =  \left(-\frac{\lambda}{3} v + \frac{\lambda}{4} v +  
	     \frac{\lambda}{4} \sigma\right) (v + \sigma)^3 =  
\\
&=& \frac{\lambda}{4} \left(\sigma-\frac{v}{3}\right) (v + \sigma)^3 =
            \frac{\lambda}{4}
	    \left(\sigma-\frac{v}{3}\right) 
	    (v^3 + \sigma^3 + 3v^2\sigma+ 3v\sigma^2) =
\\
&=& \frac{\lambda}{4} 
	    \left(\sigma(v^3 + \sigma^3 + 3v^2\sigma+ 3v\sigma^2) 
	    -\frac{v}{3} (v^3 + \sigma^3 + 3v^2\sigma+ 3v\sigma^2) 
	    \right)  =
\\
&=& \frac{\lambda}{4}\left( 
	     v^3\sigma + \sigma^4 + 3v^2\sigma^2 + 3v\sigma^3 
	    -\frac{v^4}{3}- \frac{v}{3}\sigma^3
	    - v^3 \sigma- v^2 \sigma^2\right) =
\\
&=& \frac{\lambda}{4}\left( 
	    -\frac{v^4}{3} + \sigma^4 
	    +(v^3\sigma-v^3\sigma)
	     + (3v^2\sigma^2 - v^2\sigma^2)
	     + (3 - \frac{1}{3}) v\sigma^3 \right) =
\\
&=& \frac{\lambda}{4}\left( -\frac{v^4}{3} + \sigma^4 
	     + 2v^2\sigma^2 
	     +\frac{8}{3} v\sigma^3 \right) 
	  =  
	     \frac{\lambda v^2}{2}\sigma^2 
	     +\frac{2\lambda v}{3} \sigma^3
	     + \frac{\lambda}{4}\sigma^4 	    
	     -\frac{\lambda v^4}{12}.
\end{eqnarray*}
     Finally, in terms of $\sigma$ the Lagrangian 
(\ref{Lagr-scalar-phi3}) becomes 
\begin{eqnarray} \nonumber
{\cal L} &=& \frac{1}{2} \partial_\mu \sigma \, \partial^\mu \sigma 
            - V(\sigma) = 
       \label{Phi3-sigma}
            \frac{1}{2} \partial_\mu\sigma \, \partial^\mu\sigma 
	     -\frac{\lambda v^2}{2}\sigma^2 
	     -\frac{2\lambda v}{3} \sigma^3
	     - \frac{\lambda}{4}\sigma^4 	    
	     +\frac{\lambda v^4}{12}. 
\end{eqnarray}
      This is the theory of a scalar field of mass 
      $m^2=\lambda v^2 = - v \kappa> 0$, 
      with $\sigma^3$ and $\sigma^4$ being self-interactions. 
      Note here 
      $m^2=\lambda v^2 = \frac12 m_{\kappa=0}^2$ (standard Higgs
      mechanism). 
     This is, perhaps, due to our assumption $\mu^2=0$.  
      Since the cubic terms in the initial Lagrangian,
      no any reflexion symmetry was broken.
      Therefore, one obtains a nonzero vev for the initial 
      massless scalar field $\phi$, 
      one got a mass for the new scalar field $\sigma$ 
      without any spontaneously broken symmetry.

      Now one should prove that the $\sigma$ cubic term does not
      spoil the zero-vev status of the $\sigma$ field.
      On this way one could obtain constraints on the 
      term $ -\frac{2\lambda v}{3} \sigma^3$.

     Indeed, now again one has a $\sigma^3$ term in the potential
$$
V(\sigma) =  \frac{\lambda v^2}{2}\sigma^2 
	     +\frac{2\lambda v}{3} \sigma^3
	     + \frac{\lambda}{4}\sigma^4 	    
	     -\frac{\lambda v^4}{12}
= 
\frac{\lambda}{4}\left( 2v^2 \sigma^2 
	     +\frac{8}{3} v\sigma^3
	     + \sigma^4 	    
	     - \frac{v^4}{3} \right).
$$
     The goal is to avoid any minimum at $\sigma\ne 0$.  
     Applying now the extremum condition 
(\ref{Phi3-minimum}) to this potential one can obtain 
     it in the form
\begin{equation}\label{Sigma3-minimum}
\frac{\partial V}{\partial \sigma} =
               \lambda v^2 \sigma + 2\lambda v \sigma^2 + \lambda
	       \sigma^3 
	       = \lambda \sigma (\sigma^2 + 2 v \sigma+ v^2 ) 
               = \lambda \sigma (\sigma + v)^2 
	     = 0.
\end{equation}
         There are two solutions of the equation:
$\sigma = 0$ and $\sigma = - v$. 
     Substituting both into the potential above
     one finds 
$$V(\sigma=0) = -\frac{\lambda v^4}{12} < 0 
\qquad {\rm and } \qquad 
V(\sigma=-v) = 0 .
$$
         Therefore, always a true minimum is at $\sigma=0$.

	 Finally, it looks like that in the SM with 
	 the complex SU(2) doublet of scalar fields $\Phi$ 
	 given by 
eq.~(\ref{Phi-def}) it is also 
         possible to use the ``$\phi^3$'' term
         if the Higgs potential 
(\ref{SU2-ScalarL}) is taken, say, in the form
$$ 
V(\Phi)= {\mu^2}\,|\Phi|^2 + \frac{2\sqrt{2}}{3}\kappa\,|\Phi|^3
+ {\lambda}\,|\Phi|^4,
\qquad {\rm where} \qquad
|\Phi| = \sqrt{\Phi^\dagger \Phi}. 
$$ 

\subsection{Some relations}
\label{App-proves}
        Consider in detail transformaton of the {\em Abelian Largangian}
       from section 
\ref{HM-CSF_A}, 
eq.~(\ref{Abelian-Lagrangian-1}) 
\begin{eqnarray} \label{Ab-Lag} 
{\cal L} &=&  -\frac{1}{4}F_{\mu \nu}F^{\mu \nu}
           + (\partial_\mu +ie A_\mu)\phi^* 
             (\partial^\mu -ie A^\mu)\phi 
	   - V(\phi) \\
\label{V-phi4}
&{\rm with}&~V(\phi) ~=~ \mu^2 \phi^* \phi+\lambda(\phi^* \phi)^2
\end{eqnarray}
       after substitution in it of the 
       complex scalar field in the form    
$$\displaystyle 
       \phi(x) = \frac{1}{\sqrt{2}}\left(v+\eta(x)+i\xi(x)\right)
      \equiv  \frac{1}{\sqrt{2}} (v + \phi_1 +i \phi_2)   .
$$
      The product of the covariant derivatives from 
(\ref{Ab-Lag}) becomes
\begin{eqnarray}\nonumber
(D_\mu \phi)^* (D^\mu \phi) 
&\equiv&
       (\partial_\mu +ie A_\mu)\phi^* \, (\partial^\mu -ie A^\mu)\phi 
     = (\partial_\mu \phi^* + ie A_\mu \phi^*) \,
       (\partial^\mu \phi   - ie A^\mu \phi )
\\&=&    (\partial_\mu \phi)^* (\partial^\mu \phi) 
       - ie (\partial_\mu \phi)^*  A^\mu \phi
       + ie (\partial^\mu \phi)    A_\mu \phi^* 
       + e^2 A_\mu A^\mu \, \phi^* \phi. 
\label{D_mu}
\end{eqnarray}
     Furthermore 
\begin{eqnarray*}
(\partial_\mu \phi)^* &=&  
     \left({\partial_\mu
     \frac{1}{\sqrt{2}}\left(v+\eta+i\xi\right)}\right)^* =
     \frac{1}{\sqrt{2}}\left(\partial_\mu \eta -i \partial_\mu \xi\right) ,
\\
(\partial^\mu \phi) &=&
     \left({\partial ^\mu 
     \frac{1}{\sqrt{2}}\left(v+\eta+i\xi\right)}\right) = 
     \frac{1}{\sqrt{2}}\left(\partial^\mu \eta +i\partial ^\mu \xi\right) ,
\\
(\partial_\mu \phi)^* (\partial^\mu \phi) 
&=& 
     \frac{1}{2}\left(
       \partial_\mu \eta \ \partial^\mu \eta
     +i\partial_\mu \eta \ \partial ^\mu \xi  
     -i\partial_\mu \xi  \ \partial^\mu \eta 
     + \partial_\mu \xi  \ \partial ^\mu \xi \right)
     =\frac{1}{2} \partial_\mu \eta \ \partial^\mu \eta
     +\frac{1}{2} \partial_\mu \xi  \ \partial ^\mu \xi ,
\\
(\partial_\mu \phi)^* \phi 
&=&  
     \frac{1}{{2}}\left(\partial_\mu \eta -i \partial_\mu \xi\right)
     \left(v+\eta+i\xi\right) 
 =
     \frac{v}{{2}}\left(\partial_\mu \eta -i \partial_\mu \xi\right)
    +\frac{1}{{2}}\left(\partial_\mu \eta -i \partial_\mu \xi\right)
     \left(\eta+i\xi\right) = \\
&=&
     \frac{v}{{2}}\left(\partial_\mu\eta-i\partial_\mu\xi\right)
    +\frac{1}{{2}}
     \left(\eta\partial_\mu\eta
         + i\xi\partial_\mu\eta
          -i\eta\partial_\mu\xi
         +   \xi\partial_\mu\xi
	\right),
\\
(\partial^\mu \phi) \phi^*
&=&
     \frac{1}{{2}}\left(\partial^\mu \eta +i\partial ^\mu \xi\right)
     \left(v+\eta-i\xi\right)
= 
      \frac{v}{{2}}\left(\partial^\mu \eta +i\partial ^\mu \xi\right)
    + \frac{1}{{2}}\left(\partial^\mu \eta +i\partial ^\mu \xi\right)
       (\eta-i\xi) =\\
&=&
      \frac{v}{{2}}\left(\partial^\mu \eta +i\partial ^\mu \xi\right)
    + \frac{1}{{2}}
       \left(\eta\partial^\mu\eta
            -i\xi\partial^\mu\eta
            +i\eta\partial^\mu\xi         
            +\xi\partial^\mu\xi
       \right),
\\
\phi^* \phi 
 &=&\frac{1}{2}\left(v+\eta-i\xi\right)\left(v+\eta+i\xi\right)
  = \frac{1}{2}\left( (v+\eta)^2 + \xi^2\right)
  = \frac{1}{2}\left( v^2 +2 v \eta + \eta^2 + \xi^2\right),
\\
(\phi^*\phi)^2  &=& \frac{1}{4}
        \left(v^2 +2 v \eta + \eta^2 + \xi^2\right)
        \left(v^2 +2 v \eta + \eta^2 + \xi^2\right) =
\\
&=& \frac{v^2}{4}  \left(v^2 + 2v\eta + \eta^2 + \xi^2\right) 
  + \frac{v\eta}{2}\left(v^2 + 2v\eta + \eta^2 + \xi^2\right) 
\\
&+& \frac{\eta^2}{4}\left(v^2 + 2v\eta + \eta^2 + \xi^2\right) 
  + \frac{\xi^2}{4}\left(v^2 + 2v\eta + \eta^2 + \xi^2\right) 
\\
     &=&  
      \frac{1}{4} 
       \left(
             v^4 
	   + 2v^3\eta
	   + v^2\eta^2
	   + v^2\xi^2
           + 2v^3\eta
	   + 4v^2\eta^2
	   + 2v\eta^3
	   + 2v\eta\xi^2
	   \right.
\\&+&\left. 
            v^2\eta^2
	   + 2v \eta^3
	   + \eta^4
	   + \xi^2\eta^2
           + v^2\xi^2
	   + 2v\eta\xi^2
	   + \eta^2\xi^2
	   + \xi^4
       \right) \\
&=&\frac{1}{4} 
       \left(v^4 + \eta^4 + \xi^4 + 4v^3\eta
	     + 6v^2\eta^2 + 2v^2\xi^2
	     + 4v\eta^3 + 4v\eta\xi^2 +2\eta^2\xi^2 \right).
\end{eqnarray*}
      Substituting these expansions in the derivative product 
(\ref{D_mu}) and the potential
(\ref{V-phi4})  one obtians
\begin{eqnarray*}
(D_\mu \phi)^* (D^\mu \phi) 
&=&    (\partial_\mu \phi)^* (\partial^\mu \phi)
       + e^2 A_\mu A^\mu \, \phi^* \phi  
       - ie A^\mu (\partial_\mu \phi)^*\phi  
       + ie A_\mu (\partial^\mu \phi) \phi^*    
       = \\
&=&   
       \frac{1}{2} \partial_\mu \eta \ \partial^\mu \eta
      +\frac{1}{2} \partial_\mu \xi  \ \partial ^\mu \xi
      +\frac{e^2}{2} 
         A_\mu A^\mu (v^2 +2 v \eta +\eta^2 + \xi^2)
\\ &&
    -\frac{iev}{2} A^\mu (\partial_\mu\eta-i\partial_\mu\xi)
    -\frac{ie }{2} A^\mu 
     (\eta\partial_\mu\eta
      +i\xi\partial_\mu\eta
     -i\eta\partial_\mu\xi
      +\xi\partial_\mu\xi) 
\\ &&
    +\frac{iev}{2} A_\mu(\partial^\mu\eta+i\partial ^\mu\xi)
    +\frac{ie}{{2}} A_\mu
      (\eta\partial^\mu\eta
      -i\xi\partial^\mu\eta
      +i\eta\partial^\mu\xi         
       +\xi\partial^\mu\xi) =
\\
&=&    
       \frac{1}{2} \partial_\mu \eta \ \partial^\mu \eta
      +\frac{1}{2} \partial_\mu \xi  \ \partial ^\mu \xi
      +\frac{e^2v^2}{2} A_\mu A^\mu 
      +\frac{e^2}{2} A_\mu A^\mu (2 v \eta +\eta^2 + \xi^2)
    - {ev} A^\mu \partial_\mu\xi 
\\ &&
    -\frac{ie}{2} 
    \left( A^\mu 
     (\eta\partial_\mu\eta+i\xi\partial_\mu\eta
     -i\eta\partial_\mu\xi+\xi\partial_\mu\xi) 
    - A_\mu
      (\eta\partial^\mu\eta-i\xi\partial^\mu\eta
      +i\eta\partial^\mu\xi+\xi\partial^\mu\xi) \right)
\\
(D_\mu \phi)^* (D^\mu \phi) 
&=&    
       \frac{1}{2} \partial_\mu \eta \ \partial^\mu \eta
      +\frac{1}{2} \partial_\mu \xi  \ \partial ^\mu \xi
      +\frac{e^2v^2}{2} A_\mu A^\mu 
      +\frac{e^2}{2} A_\mu A^\mu (2 v \eta +\eta^2 + \xi^2)
    - {ev} A^\mu \partial_\mu\xi 
\\ &+&
         e A^\mu \xi\partial_\mu\eta 
       - e A^\mu \eta\partial_\mu\xi .
\\
V(\phi) &=& \mu^2 \phi^* \phi+\lambda(\phi^* \phi)^2
        =  \frac{\mu^2}{2} \left(v^2+2v\eta+\eta^2+\xi^2\right)
\\
        &+&\frac{\lambda}{4}
       \left(v^4 + \eta^4 + \xi^4 + 4v^3\eta
	     + 6v^2\eta^2 + 2v^2\xi^2
	     + 4v\eta^3 + 4v\eta\xi^2 +2\eta^2\xi^2 \right).
\end{eqnarray*}
     With the minimun realtion $\mu^2=-v^2\lambda$\/ 
     the potential becomes
\begin{eqnarray*}
V(\phi) 
     &=&-\frac{v^2\lambda}{2} \left(v^2+2v\eta+\eta^2+\xi^2\right)
\\
     &+&\frac{\lambda}{4}
       \left(v^4 + \eta^4 + \xi^4 + 4v^3\eta
	     + 6v^2\eta^2 + 2v^2\xi^2
	     + 4v\eta^3 + 4v\eta\xi^2 +2\eta^2\xi^2 \right) =
\\
     &=& 
	-\frac{v^4\lambda}{2} 
	-{v^3\lambda}\eta
	-\frac{v^2\lambda}{2}\eta^2
	-\frac{v^2\lambda}{2}\xi^2 
\\&&
       +\frac{v^4\lambda}{4} 
       + \frac{\lambda}{4}\eta^4 
       + \frac{\lambda}{4}\xi^4 
       + {v^3\lambda}\eta
       + \frac{3v^2\lambda}{2}\eta^2 
       + \frac{v^2\lambda}{2}\xi^2
       + {v\lambda}\eta^3 
       + {v\lambda}\eta\xi^2 
       + \frac{\lambda}{2}\eta^2\xi^2 
\\
V(\phi) &=& 
	-\frac{v^4\lambda}{4}
	+{v^2\lambda}\eta^2 
       + \frac{\lambda}{4}\eta^4 
       + \frac{\lambda}{4}\xi^4 
       + {v\lambda}\eta^3 
       + {v\lambda}\eta\xi^2 
       + \frac{\lambda}{2}\eta^2\xi^2.
\end{eqnarray*}
       Collecting now all terms together one can obtain 
       the Abelian Lagrangian 
(\ref{Ab-Lag}) in the form 
\begin{eqnarray} \nonumber
{\cal L} &=&-\frac{1}{4}F_{\mu \nu}F^{\mu \nu}
            + (D_\mu \phi)^* (D^\mu \phi) 
	    - V(\phi) = 
\\\label{Ab-Lag-fino} 
	&=& -\frac{1}{4}F_{\mu \nu}F^{\mu \nu} 
            +\frac{1}{2}\partial_\mu\eta\,\partial^\mu\eta
            +\frac{1}{2}\partial_\mu\xi \,\partial ^\mu\xi
\\&& \nonumber 
      +\frac{e^2v^2}{2} A_\mu A^\mu 
      +\frac{e^2}{2} A_\mu A^\mu (2 v \eta +\eta^2 + \xi^2)
    - {ev} A^\mu \partial_\mu\xi 
       +  e A^\mu \xi\partial_\mu\eta 
       - e A^\mu \eta\partial_\mu\xi
\\ && \nonumber 
       - {v^2\lambda}\eta^2 
       - \frac{\lambda}{4}(\eta^2 + \xi^2)^2
       - {v\lambda}\eta (\eta^2 +\xi^2) 
       +\frac{v^4\lambda}{4}.
\end{eqnarray}
        With $\eta=\phi_1$ and $\xi=\phi_2$
	the relevant Lagrangian from 
\cite{Gaillard:1977wu} is 
\begin{eqnarray*} 
{\cal L}&\equiv& -\frac{1}{4}F_{\mu \nu}F^{\mu \nu}
           +(\partial^\mu +ieA^\mu)\phi^*(\partial_\mu -ieA_\mu)\phi 
           -\mu^2\phi^*\phi-\lambda(\phi^*\phi)^2 = 
\\ 
      &=& 
          -\frac{1}{4}F_{\mu \nu}F^{\mu \nu} 
          +\frac{1}{2}(\partial_\mu \phi_1)^2 
          +\frac{1}{2}(\partial_\mu \phi_2)^2
          -\phi_1 (\mu^2+\lambda v^2)
          -\frac{1}{2} \phi_2^2 (\mu^2 +  v^2 \lambda)          
\\ &&
          +\frac{e^2 v^2 }{2} A_\mu A^\mu
 	  + \frac{e^2}{2} A_\mu A^\mu   (\phi_1^2 + \phi_2^2)
	  + e^2 v 
	  A_\mu A^\mu \phi_1
          - e v A_\mu \partial^\mu \phi_2
	  + e A_\mu (\partial^\mu \phi_1) \phi_2 
	  - e A_\mu (\partial^\mu \phi_2) \phi_1
\\ &&
          -\frac{1}{2} \phi_1^2 (\mu^2 + 3v^2 \lambda) 
	  - \frac{\lambda}{4}(\phi_1^2 +\phi_2^2)^2
          -\lambda v \phi_1 (\phi_1^2 + \phi_2^2)  
          +\frac{v^4\lambda}{4}. 
\end{eqnarray*}

       In the unitary gauge  
(\ref{Ab-unitary-gauge}) Lagrangian
(\ref{local-U1}) or 
(\ref{Ab-Lag-fino}) with $\mu^2=-v^2\lambda$ 
transforms as follows:
\begin{eqnarray*} 
{\cal L}
&=& \frac{1}{2}(\partial_\mu +ieA_\mu)(v +\eta) 
              (\partial^\mu -ieA^\mu)(v +\eta) 
     -\frac{\mu^2}{2}(v +\eta)^2 
     -\frac{\lambda}{4}(v +\eta)^4 -\frac{F_{\mu\nu}F^{\mu\nu}}{4} =
\\
&=& \frac{1}{2}
      (\partial_\mu\eta +ieA_\mu (v +\eta))
      (\partial^\mu\eta -ieA^\mu (v +\eta)) 
      -\frac{F_{\mu\nu}F^{\mu\nu}}{4}
\\&&
     -\frac{v^2\lambda}{2}(-v^2 -2v\eta -\eta^2)
     -\frac{\lambda}{4} 
    (v^4 + 4v^3\eta +6 v^2\eta^2 +4v\eta^3 + \eta^4) =
\\
&=& \frac{1}{2}
      (
      \partial_\mu\eta\partial^\mu\eta 
     -ie \partial_\mu\eta A^\mu(v +\eta)
     +ie A_\mu\partial^\mu\eta (v +\eta)
      )
     +\frac{1}{2}
     e^2A_\mu A^\mu (v^2 +2 v \eta + \eta^2) 
\\&&
     -\frac{F_{\mu\nu}F^{\mu\nu}}{4}
     -\frac{\lambda}{4}(-2v^4 -4v^3\eta -2v^2\eta^2
     +v^4 + 4v^3\eta +6 v^2\eta^2 +4v\eta^3 + \eta^4) \to
\\
{\cal L}
&=&  \frac{1}{2}
      \partial_\mu\eta\partial^\mu\eta 
     -\frac{2v^2\lambda}{2}\eta^2 
     -\frac{F_{\mu\nu}F^{\mu\nu}}{4}
     +\frac{e^2v^2}{2}A_\mu A^\mu 
     +\frac{e^2}{2}A_\mu A^\mu (2 v \eta + \eta^2) 
     +\frac{v^4\lambda}{4}
     -{\lambda}v\eta^3 
     -\frac{\lambda}{4}\eta^4.
\end{eqnarray*}

       If one ignores the vector field $V_\mu$ 
       (consider the pure complex scalar field, see 
section~\ref{HM-CSF} and
(\ref{complex-1})) 
       and introduces the Higgs mass
$ m_H = \sqrt{2 \lambda v^2} \equiv \sqrt{2}|\mu|$\/ 
       the Lagrangian takes the form
\begin{eqnarray*}
{\cal L}
     &=& 
      \frac{1}{2}\partial_\mu\eta\,\partial^\mu\eta
     +\frac{1}{2}\partial_\mu\xi  \,\partial^\mu\xi
     -\frac{\lambda}{4}\eta^4
     -\frac{\lambda}{4}\xi^4
     -\frac{\lambda}{2}\eta^2\xi^2
     + \frac{v^4\lambda}{4} 
     - \frac{m^2_H}{2}\eta^2  
     - v\lambda\eta^3
     - v\lambda\eta\xi^2.
\end{eqnarray*}

    Consider transformation of the product of the covariant
    derivatives $|D_\mu\Phi|^2$ 
(\ref{SM-Dmu}) 
    in the Standard Model (see section \ref{HM-SM}).
    First, one has to insert  
    the explicit forms of Pauli matrices (\ref{Pauli-matrices}) 
    into $D_\mu\Phi$ and then contract $(D_\mu\Phi)^\dagger$ and 
    $|D_\mu\Phi|^2$: 
{\small
\begin{eqnarray*}
D_\mu\Phi &=& \left(\partial_\mu - ig_2 \frac{\tau_a}{2} W_\mu^a 
               - i g_1 \frac{Y_H}{2}B_\mu \right)\Phi =
\\&=&\left(
	   \left(\begin{array}{cc}1 & 0\\0 & 1 \end{array}\right)
	   \left(\partial_\mu-\frac{ig_1Y_H}{2}B_\mu\right)
          -\frac{ig_2}{2} 
	   \left(
	   \left(\begin{array}{cc}0 & 1\\1 & 0\end{array}\right) W_\mu^1
	  +\left(\begin{array}{cc}0 &-i\\i & 0\end{array}\right) W_\mu^2 
	  +\left(\begin{array}{cc}1 & 0\\0 &-1\end{array}\right) W_\mu^3 
          \right) 
	     \right) \Phi \\ 
     &=&
     \left(\begin{array}{cc}
	\partial_\mu
	 -\frac{i}{2}\left({g_2}W_\mu^3+{g_1Y_H}B_\mu\right) 
	&-\frac{ig_2}{2}(W_\mu^1-i W_\mu^2) 
       \\-\frac{ig_2}{2}(W_\mu^1+i W_\mu^2) 
	& \partial_\mu
	 +\frac{i}{2}\left({g_2}W_\mu^3-{g_1Y_H}B_\mu\right) 
                 \end{array}\right)
\frac{1}{\sqrt{2}}\left(\begin{array}{c}0\\v+h(x)\end{array}\right)=\\    
 &=&
     \frac{1}{\sqrt{2}}
     \left(\begin{array}{c}
     -\frac{ig_2}{2}(W_\mu^1-i W_\mu^2)(v+h(x))
     \\\partial_\mu h(x)
       +\frac{i}{2}\left({g_2}W_\mu^3-{g_1Y_H}B_\mu\right)(v+h(x))
     \end{array}\right); 
\\  
(D^\mu\Phi)^\dagger
&=&  \Phi^\dagger
     \left(\partial^\mu-ig_2\frac{\tau_a}{2}W^\mu_a
           -ig_1\frac{Y_H}{2}B^\mu\right)^\dagger 
 =  \frac{1}{\sqrt{2}}
     \left(\begin{array}{c}
     -\frac{ig_2}{2}(W^\mu_1-i W^\mu_2)(v+h(x))
     \\  \partial^\mu h(x)
	 +\frac{i}{2}\left({g_2}W^\mu_3-{g_1Y_H}B^\mu\right)(v+h(x))
     \end{array}\right)^\dagger  
\\ &=&
    \frac{1}{\sqrt{2}} \left(
     \frac{ig_2}{2}(W_\mu^1+i W_\mu^2)(v+h(x)), \
     \partial_\mu h(x)
      -\frac{i}{2}\left({g_2}W_\mu^3-{g_1Y_H}B_\mu\right)(v+h(x))
      \right);
\\
|D_\mu\Phi|^2 
  &\equiv& 
(D^\mu\Phi)^\dagger(D_\mu\Phi) =
\left|\left(\partial_\mu - ig_2 \frac{\tau_a}{2} W_\mu^a 
               - i g_1 \frac{Y_H}{2}B_\mu \right)\Phi \right|^2 =
\\ &=&  
     \Phi^\dagger
     \left(\partial^\mu-ig_2\frac{\tau_a}{2}W^\mu_a
             -ig_1\frac{Y_H}{2}B^\mu\right)^\dagger 
      \left(\partial_\mu - ig_2 \frac{\tau_a}{2} W_\mu^a 
              -ig_1\frac{Y_H}{2}B_\mu\right)\Phi =
\\
&=& \frac{1}{2}\left| 
      \left(\begin{array}{cc} 
       \partial_\mu-\frac{i}{2}(g_2 W_\mu^3+g_1 Y_H B_\mu) 
     &-\frac{ig_2}{2}(W_\mu^1-iW^2_\mu) 
    \\-\frac{ig_2}{2}(W_\mu^1+iW^2_\mu) 
     & \partial_\mu+\frac{i}{2}(g_2 W_\mu^3-g_1 Y_H B_\mu) 
            \end{array}\right) 
         \left(\begin{array}{c}0\\v+h(x)\end{array}\right) 
	 \right|^2 
\\
&=& 
     \frac{1}{\sqrt{2}}\left(
     \frac{ig_2}{2}(W^\mu_1+i W^\mu_2)(v+h),\
     \partial^\mu h(x)
     -\frac{i}{2}\left({g_2}W^\mu_3-{g_1Y_H}B^\mu\right)(v+h)
      \right) 
\\&&\times
     \frac{1}{\sqrt{2}}
     \left(\begin{array}{c}
     -\frac{ig_2}{2}(W_\mu^1-i W_\mu^2) (v+h)
     \\ \partial_\mu h(x)
	+\frac{i}{2}\left({g_2}W_\mu^3-{g_1Y_H}B_\mu\right)(v+h)
     \end{array}\right)=
\\
&=&
     \frac{1}{2}
     \left(-\frac{ig_2}{2}(W^\mu_1+i W^\mu_2)(v+h)
            \frac{ig_2}{2}(W_\mu^1-i W_\mu^2)(v+h) \right. 	 
\\ 
&+& \left.\left(
      \partial^\mu h -\frac{i}{2}({g_2}W^\mu_3-{g_1Y_H}B^\mu)(v+h)
      \right) \left(
      \partial_\mu h +\frac{i}{2}({g_2}W_\mu^3-{g_1Y_H}B_\mu)(v+h)
      \right) \right) =
\\&=&
     \frac{g^2_2}{8}(v+h)^2(W^\mu_1+i W^\mu_2)(W_\mu^1-i W_\mu^2)
     +\frac{1}{2}
     \partial^\mu h(x)\partial_\mu h(x)
     +\frac{1}{8}(v+h)^2\left({g_2}W^\mu_3-{g_1Y_H}B^\mu\right)^2,
\\
|D_\mu\Phi|^2 
&=&
  \frac{1}{2}(\partial_\mu h)^2 
+ \frac{g_2^2}{8}(v+h)^2|W_\mu^1+iW_\mu^2|^2
+ \frac{1}{8}(v+h)^2(g_2 W_\mu^3- g_1 Y_H B_\mu)^2.
\end{eqnarray*}
}

\providecommand{\href}[2]{#2}\begingroup\raggedright\endgroup

\end{document}